\documentclass[final,3p,times]{elsarticle}
\usepackage{amssymb}
\usepackage{amsmath}
\usepackage{url}
\usepackage{diagbox}
\usepackage{longtable}
\usepackage{hhline}
\usepackage{multirow}
\usepackage{rotating}
\usepackage{colortbl}
\usepackage{todonotes}
\usepackage{graphicx}
\usepackage{subcaption}
\usepackage{algorithm}
\usepackage{algpseudocode}
\usepackage{tikz}
\usetikzlibrary{shapes.geometric, arrows, positioning, fit}

\usepackage{tabularx}
\usepackage{xcolor}
\definecolor{rev2}{rgb}{0.75,0.55,0.0}
\journal{International Journal of Human-Computer Studies}

\begin{document}

\begin{frontmatter}

% \title{Quantifying the Impact of Motion on 2D Gaze Estimation in Real-World Mobile Interactions} %% Article title

\title{Quantify2DGaze: Quantifying Motion-Related Degradation in 2D Gaze Estimation during Handheld Mobile Interaction}

\author[label1]{Yaxiong Lei}
\ead{yaxiong.lei@st-andrews.ac.uk}

\author[label1]{Yuheng Wang}
\ead{yw99@st-andrews.ac.uk}

\author[label1]{Mingyue Zhao}
\ead{mz201@st-andrews.ac.uk}

\author[label2]{Shijing He}
\ead{shijing.he@kcl.ac.uk}

\author[label1]{Fergus Buchanan}
\ead{fb206@st-andrews.ac.uk}

\author[label3]{Yusuke Sugano}
\ead{sugano@iis.u-tokyo.ac.jp}

\author[label4]{Mohamed Khamis}
\ead{mohamed.khamis@glasgow.ac.uk}

\author[label1]{Juan Ye}
\ead{juan.ye@st-andrews.ac.uk}

\affiliation[label1]{organization={University of St Andrews},
            city={St Andrews},
            postcode={KY16 9SX},
            country={UK}}

\affiliation[label2]{organization={King's College London},
            city={London},
            postcode={WC2R 2LS},
            country={UK}}

\affiliation[label3]{organization={University of Tokyo},
            city={Tokyo},
            postcode={113-0033},
            country={JP}}

\affiliation[label4]{organization={University of Glasgow},
            city={Glasgow},
            postcode={	G12 8QQ},
            country={UK}}

%% Abstract
% \begin{abstract}
% Mobile gaze tracking involves inferring a user's gaze point or direction on a mobile device's screen from facial images captured by the device's front camera. While this technology inspires an increasing number of gaze-interaction applications, achieving consistent accuracy remains challenging due to dynamic user-device spatial relationships and varied motion conditions inherent in mobile contexts. This paper provides empirical evidence on how user mobility and behaviour affect mobile gaze tracking accuracy. We conduct two user studies collecting behaviour and gaze data under various motion conditions—from lying to maze navigation—and during different interaction tasks. Quantitative analysis has revealed behavioural regularities among daily tasks and identified head distance, head pose, and device orientation as key factors affecting accuracy, with errors increasing by up to 48.91\% in dynamic conditions compared to static ones. These findings highlight the need for more robust, adaptive eye-tracking systems that account for head movements and device deflection to maintain accuracy across diverse mobile context.
% \end{abstract}

\begin{abstract}
Mobile gaze tracking on handheld devices commonly relies on \emph{appearance-based 2D gaze estimation} from the device's \emph{front-facing RGB camera}. Although this approach makes eye tracking more accessible by avoiding dedicated infrared hardware, maintaining accuracy remains difficult because the spatial relationship among the user's face, the camera, and the screen changes continuously during handheld use. This paper examines how motion-related geometry change affects a representative front-camera RGB 2D gaze-estimation pipeline. We report two complementary within-subject user studies (\(n=10\)) that capture handheld behaviour, calibration data, and gaze-estimation error during everyday-style mobile tasks across conditions ranging from lying and sitting to walking and maze navigation.

Rather than treating motion as a single binary factor, we analyse within-task versus task-switch behaviour, calibration-test mismatch, and condition-specific motion signatures. Behavioural results show that head-to-camera distance is comparatively stable during sustained within-task use, while task switches introduce much larger geometry fluctuations, with head-distance standard deviations of approximately 7.42--10.10\,cm across conditions. In the gaze-evaluation study, one-off calibration reduced average RMSE from 4.49\,cm to 3.44\,cm; under this regime, the global effect of motion condition on RMSE was not statistically reliable, whereas clearer patterns appeared in task-specific and motion-specific calibration analyses, where better calibration-test match reduced error.

% \textcolor{blue}{Rather than treating motion as a single binary factor, we analyse within-task versus switch-task behaviour, calibration-test mismatches, and condition-specific motion signatures. The results show that within a task, head-to-camera distance is comparatively stable, whereas task switches are associated with markedly larger fluctuations; these transition periods coincide with larger geometric changes that are poorly handled by a one-off calibration. In our studies, error under dynamic conditions was higher than under static conditions by up to \textcolor{rev2}{23.5\%} depending on calibration regime, and factor analysis consistently ranked head-to-camera distance, device-orientation change, and head movement among the strongest contributors to error. We further show that these contributions differ by condition, suggesting that aggregated coefficients should be interpreted only with respect to the present experimental design rather than as general properties of mobile use.}

Factor analyses repeatedly associated head-to-camera distance, head movement, and device-orientation change with gaze-estimation error, with their relative importance varying across motion conditions. Taken together, the findings provide a condition-sensitive account of motion-related degradation in handheld 2D gaze estimation and suggest design directions for motion-aware calibration, geometry-aware confidence estimation, and lightweight 2D--3D hybrid modelling. Because the study uses a small sample and semi-controlled laboratory tasks, the reported effect sizes should be interpreted as exploratory, design-oriented evidence rather than population-level estimates.

\end{abstract}

%% Keywords

\end{frontmatter}




\section{Introduction}

Eye tracking on handheld mobile devices has recently emerged as an important direction in human-computer interaction research. Unlike traditional infrared eye trackers or near-eye systems embedded in virtual reality (VR) and augmented reality (AR) headsets, handheld mobile eye tracking typically relies on the device's front-facing RGB camera to infer where a user is looking~\cite{lei2023end, blignaut2016idiosyncratic, menges2019improving}. This is referred to as \emph{appearance-based gaze estimation}. 

Compared with specialised eye trackers, appearance-based methods offer substantially greater scalability and accessibility, making mobile eye tracking feasible for everyday devices. This capability is increasingly important because gaze interaction on mobile devices is becoming relevant to a growing range of applications, including attentive user interfaces, accessibility support, reading analysis, authentication, adaptive content delivery, and context-aware interaction~\cite{khamis2018understanding, lei2023DynamicRead}. At the same time, major device manufacturers and technology companies are investing heavily in gaze-enabled interaction across mobile, wearable, and spatial computing ecosystems. As eye tracking becomes more deeply integrated into consumer devices, improving the robustness of appearance-based mobile gaze estimation has become both a practical and scientific challenge.

Appearance-based gaze estimation can be broadly categorised by output representations. \emph{2D gaze estimation} predicts gaze coordinates directly on the device screen, whereas \emph{3D gaze estimation} predicts gaze vectors or directions in space~\cite{sugano2014learning, zhang18revisiting, cheng2024benchmark}. Among these approaches, 2D gaze estimation has become especially popular for handheld mobile interaction because it naturally supports screen-based applications and can be deployed using commodity front-facing cameras without additional hardware~\cite{krafka2016eye, arakawa2022rgbdgaze, huynh2021imon, valliappan2020accelerating}. .

A central challenge in gaze estimation is \emph{calibration}. Accurate eye tracking generally requires a calibration procedure that establishes the relationship between eye appearance, device geometry, and gaze output~\cite{duchowski2017eye,tobii2022calib}. In many conventional eye-tracking systems, calibration assumes that the spatial relationship between the user, camera, and display remains relatively stable after calibration. However, this assumption is considerably weaker for handheld mobile devices. During ordinary mobile interaction, users naturally change posture, viewing distance, head pose, and device orientation while sitting, walking, switching tasks, or interacting in different environments~\cite{huang2019saccalib, martschinke2019gaze, lei2023DynamicRead}. These changes alter the face-camera-screen geometry captured by the front camera and can substantially affect gaze-estimation performance~\cite{cheng2024benchmark, lei2023DynamicRead}.

Prior work has identified several factors that influence mobile gaze estimation, including head movement, device motion, lighting conditions, and viewing distance~\cite{khamis2018understanding, bace2020quantification, huang2017tabletgaze}. Existing studies have also explored calibration techniques, dataset construction, and model architectures for both 2D and 3D gaze estimation~\cite{krafka2016eye, balim2023efe, bace2019accurate}. However, important gaps remain. First, much of the literature studies these factors in isolation rather than examining how they interact during everyday handheld use. Second, there is still limited understanding of \emph{when} geometry changes occur during mobile interaction and how strongly those changes affect calibration stability. Third, although calibration distortion~\cite{huang2019saccalib, martschinke2019gaze, lei2023DynamicRead} is widely acknowledged as a problem, there is relatively little quantitative evidence about how different calibration and testing conditions interact under realistic mobile behaviour.

Our work aims to provide quantitative answers to these questions. More specifically, we investigate how motion-related geometry changes affect handheld 2D gaze estimation during semi-controlled mobile interaction, with particular attention to calibration-test mismatch and behavioural transitions between tasks. Rather than treating motion as a uniform source of noise, we examine when geometry changes occur, which observable motion-related signals are most associated with error, and how calibration performance varies across interaction conditions.

To address these gaps, we conduct two complementary user studies and develop an experimental framework for synchronised motion-and-gaze capture on handheld devices. Our main contributions are listed below. 

\begin{enumerate}

\item We characterise behavioural regularities during everyday-style mobile interaction, and analyse both within-task stability and task-switch transitions. Our findings show that head-to-camera distance remains comparatively stable during sustained task interaction, but fluctuates substantially during transitions between tasks.

\item We evaluate a front-camera RGB 2D gaze-estimation pipeline under different calibration and motion conditions. Our analyses show that one-off calibration improves performance but does not adequately account for later geometry variation, whereas better calibration-test alignment reduces gaze-estimation error. Across conditions, head-to-camera distance, head movement, and device-orientation change consistently emerge as key factors associated with performance degradation.

\item Based on these findings, we derive implications for motion-aware calibration strategies, adaptive confidence estimation, and geometry-aware gaze-estimation models for handheld mobile devices.

\end{enumerate}

\section{Related Work}\label{sec:relatedwork}

Our research intersects four areas: (1) gaze estimation on mobile devices, (2) eye-tracking calibration, (3) user behaviour modelling, and (4) robustness evaluation for mobile gaze estimation. This section reviews these areas to position our study as an empirical analysis of how handheld geometry, calibration, and gaze-estimation error interact within the same evaluation pipeline.

\subsection{Gaze Estimation on Mobile Devices}

Eye tracking research on mobile devices has gained significant traction in recent years, driven by improvements in mobile processing capability, camera quality, and the broader shift of everyday interaction from desktops to mobile platforms~\cite{khamis2018past, cheng2024benchmark, lei2023end, ghosh2023automatic}. Existing mobile gaze-estimation approaches can be broadly grouped into model-based and appearance-based methods.

Model-based approaches use geometric assumptions about the face, eyes, camera, and screen to estimate gaze. They can be interpretable, but some required parameters, such as eye centre, are not directly observable~\cite{hansen2009eye, kaur2022rethinking}. They are also sensitive to image quality and scene constraints~\cite{hansen2009eye, liu20203d}, which makes deployment difficult across the changing lighting, posture, and motion conditions of handheld use. As a result, relatively few mobile systems have adopted fully model-based designs, with EyeTab~\cite{wood2014eyetab} and ScreenGlint~\cite{huang2017screenglint} as notable examples.

Appearance-based approaches instead learn a mapping from facial images to gaze targets using data-driven models~\cite{zhang2015appearance, zhang2017fullface}. Because they can operate with standard RGB cameras and do not strictly require infrared illumination, they are attractive for handheld devices~\cite{zhang2015appearance, krafka2016eye}. These methods can be further categorised by output representation: \textit{2D} gaze estimation predicts screen coordinates, whereas \textit{3D} gaze estimation predicts gaze directions as vectors in space~\cite{sugano2014learning, cheng2024benchmark}.

Methods for appearance-based gaze estimation can be further distinguished by their approach to handling head pose variability. 
% 3D methods often involve head pose estimation and normalization techniques, where facial images are transformed into a normalized space to reduce variability due to head pose and other factors~\cite{zhang2018revisiting, zhang19mpiigaze}. By decoupling head pose from gaze direction, 3D methods can generalize better across different head poses and device configurations. However, applying 3D methods in mobile settings presents challenges, i.e. the data normalisation process is usually performed offline as its cost~\cite{park2019few}, which causes additional latency in mobile setting for real-time use. Recent approach, like EFE~\cite{balim2023efe}, try to reduce the pre-processing expensive cost to compute the gaze vector directly from the frame and then calculate gaze points via ray-plane intersection, leveraging the 3D gaze vector and camera parameters.
A key distinction between 2D and 3D gaze estimation is how geometry change is handled. 3D pipelines typically incorporate head-pose estimation and camera normalization so that gaze direction can be inferred in a canonical space~\cite{zhang2018revisiting, zhang2017mpiigaze}. This makes them conceptually better suited to pose variation, but such pipelines can introduce additional pre-processing and post-processing complexity and computational cost in handheld settings~\cite{park2019few, balim2023efe}. By contrast, 2D methods are operationally simpler but often rely on the model and calibrator to absorb geometry variation implicitly. Our study is motivated by the practical question of where this simplification begins to fail during handheld interaction, and which geometry-related factors should be prioritised if one wishes to make 2D pipelines more robust.

% With the release of GazeCapture~\cite{krafka2016eye}, a dataset specifically for mobile devices, and subsequent datasets such as RGBDGaze~\cite{arakawa2022rgbdgaze}, 2D gaze estimation has gradually become one of dominant algorithms behind eye tracking on mobile devices due to its simplicity and ease of deployment~\cite{lei2023end, cheng2024benchmark}. 
% Various variants are derived with different model input settings, from eyes~\cite{valliappan2020accelerating}, to various permutations of eyes, faces, landmarks, etc.~\cite{krafka2016eye, huang2022gazeattentionnet, bao2021adaptive, huynh2021imon}, and various model architectures, e.g., CNN~\cite{zhang2015appearance,zhang2017fullface, krafka2016eye}, CNN variants combined with Recurrent Neural Networks (RNNs)~\cite{palmero2019recurrent,kellnhofer2019gaze360}, and more recent transformers~\cite{cheng2022gaze} to continuously achieve better results on the datasets. 
With the release of GazeCapture~\cite{krafka2016eye} and later datasets such as RGBDGaze~\cite{arakawa2022rgbdgaze}, 2D gaze estimation has become one of the dominant approaches behind handheld eye tracking because of its simplicity and ease of deployment~\cite{lei2023end, cheng2024benchmark}. Existing work has explored different input configurations, ranging from eye crops alone~\cite{valliappan2020accelerating} to combinations of eyes, face, and landmarks~\cite{krafka2016eye, huang2022gazeattentionnet, bao2021adaptive, huynh2021imon}, as well as different model families including CNNs~\cite{zhang2015appearance, zhang2017fullface, krafka2016eye}, recurrent variants~\cite{palmero2019recurrent, kellnhofer2019gaze360}, and transformer-based architectures~\cite{cheng2022gaze}. These studies mainly advance benchmark accuracy under their respective training and evaluation protocols. However, benchmark improvements do not by themselves explain how performance degrades when handheld geometry changes during use, how calibration interacts with motion, or which motion-related factors dominate error under different usage regimes.

% These ongoing efforts continue to push the boundaries of accuracy and efficiency in mobile gaze estimation. However, the inherent challenges of mobile environments such as variable lighting conditions, diverse usage postures, and constant device movement require for robust solutions that can handle the dynamic nature of mobile interactions. Our work builds upon these foundations, focusing on the limitations of 2D gaze estimation approaches in handling head pose variability, a critical factor in real-world mobile usage scenarios.

Our study uses a representative pipeline of this kind as an empirical probe rather than proposing a new architecture, and instead asks how handheld motion, task switching, and calibration-test mismatch alter its operating geometry and the resulting error.

\subsection{Eye-tracking Calibration}
Calibration adapts an eye-tracking system to the current spatial relationship between the user, camera, and display~\cite{lei2023end, duchowski2017eye}. In common calibration procedures, the interface asks users to fixate on known points~\cite{Drewes2019timecalibration,krafka2016eye, valliappan2020accelerating, huynh2021imon} or follow a moving target~\cite{lei2023DynamicRead,Drewes2019timecalibration}, producing paired samples of gaze target and model observation. A calibrator then adjusts the output of the base gaze-estimation model for the current user and device geometry.

In mobile 2D gaze estimation, Valliappan et al.~\cite{valliappan2020accelerating} used Support Vector Regression (SVR) as a one-off calibrator and reported high accuracy under a 30-second seated calibration setting. This demonstrates the value of user-specific calibration, but also highlights an important deployment question for handheld use: a calibrator trained under one posture or task may not cover the geometry encountered later when the user changes task, hand posture, or motion condition.

Calibration has also been extensively explored for desktop and remote eye trackers~\cite{lei2023end}. Sugano et al.~\cite{Sugano08Incremental} proposed incremental learning with mouse clicks as implicit gaze labels and clustered training data by head pose to create pose-specific calibrators. Huang et al.~\cite{huang2019saccalib} used saccade linearity to correct gaze prediction errors, while Pi and Shi~\cite{pi2019task} proposed a position-dependent homography correction after observing systematic degradation from head-position changes. Li et al.~\cite{li2022calibration} used Monte Carlo analysis to estimate the variability of eye-tracking accuracy under challenging conditions and anticipate when recalibration may be needed.

These studies show that calibration can be made more adaptive, but much of the evidence comes from desktop, seated, or otherwise constrained settings. Handheld use adds a coupled face-camera-screen geometry in which the user and device move together. Our work therefore examines calibration in a handheld 2D pipeline by asking how one-off, task-specific, and motion-specific calibration regimes behave when calibration and test geometry differ.

\subsection{User Behaviour Modelling on Handheld Mobile Devices}
Handheld gaze estimation depends not only on the visual model, but also on how users hold, move, and reorient the device during interaction. Mobile devices include sensors that can capture these behaviours, making user-behaviour modelling relevant for understanding when gaze-estimation conditions change. Prior work has studied interruptions~\cite{adamczyk2004if, fogarty2005predicting}, attention switching~\cite{steil2018forecasting}, and attention allocation in everyday life~\cite{bace2020quantification}. Other studies have used device cameras or head-worn cameras to track visual attention and predict attention shifts~\cite{jiang2016vads, steil2018forecasting}.

In mobile gaze estimation, behaviour modelling is especially relevant because mobility and holding posture change the viewing geometry. Huang et al.~\cite{huang2017screenglint} used the front camera to estimate face-to-screen distance during different activities, and Lei et al.~\cite{lei2023DynamicRead} showed that mobility such as walking can reduce the usability of gaze interfaces requiring high accuracy, such as smooth pursuit. Gaze datasets such as MPIIGaze~\cite{zhang2017mpiigaze}, EyeDiap~\cite{funes2014eyediap}, TableGaze~\cite{huang2017tabletgaze}, GazeCapture~\cite{krafka2016eye}, and RGBDGaze~\cite{arakawa2022rgbdgaze} provide important resources for modelling gaze from images, but they do not fully characterise how handheld motion, task transitions, calibration events, and gaze-error outcomes interact within a single phone-use protocol.
% There are no datasets nor studies that record IMU sensor data and eye-tracking data when users are performing daily tasks on their devices. 
Several systems and datasets point toward this combined view. EyeMU~\cite{kong2021eyemu}, Gaze-in-Wild~\cite{gazeinwild2020}, gaze prediction from egocentric video and IMU~\cite{egocentricimu2021}, and IMU-assisted gaze-zone estimation in driving~\cite{driverimu2023} demonstrate the value of combining gaze-related analysis with motion or egocentric sensing. However, comparatively less is known about synchronised handheld IMU signals, front-camera observations, calibration regimes, and 2D gaze-estimation error \emph{within the same evaluation pipeline}. Our work targets this narrower gap by coupling behaviour analysis and gaze-error analysis in a single semi-controlled handheld setup.

% Our study addresses this gap by linking behavioural motion patterns with gaze-error analysis in a semi-controlled handheld setup.

% Our paper aims to bridge the gap, encouraging users to perform various types of tasks on their phones under different motion conditions. The data will help us understand their behaviours under different tasks and as well as when switching tasks. The understanding will uncover insights on how user behaviours impact gaze estimation. 

\subsection{Robustness Evaluation of Gaze Estimation}
Robustness evaluation for mobile gaze estimation asks how accuracy changes when the conditions at test time differ from those seen during training or calibration. Unlike fixed desktop or head-mounted settings, handheld systems must contend with a changing spatial relationship among the user's eyes, the device camera, and the screen. This makes it important to identify which image, pose, distance, and motion factors are associated with error.

Prior studies have identified several factors that degrade performance. Lower image resolution can reduce the information available for gaze prediction~\cite{zhang2017mpiigaze}. Lighting changes, uneven illumination, shadows, and glare can compromise prediction, especially when models are trained on limited lighting conditions~\cite{zhang2017mpiigaze, zhang2020eth}. Facial occlusion is another source of error, with partially occluded faces causing substantial performance drops~\cite{zhang2017fullface}.

Head pose and viewing distance are particularly relevant to handheld use. Prior work reports performance drops under unseen head poses~\cite{zhang2020eth}; Valliappan et al.~\cite{valliappan2020accelerating} showed that error increases as head roll, pan, and tilt deviate from the calibrated pose, and that reduced face ratio, a proxy for increased distance, also increases error. Huang et al.~\cite{huang2017screenglint} similarly reported higher error as face-to-screen distance increased.

Variable holding patterns add another layer of difficulty. Khamis et al.~\cite{khamis2018understanding} found that front cameras capture users' full faces only 29\% of the time during unconstrained use, due to the diversity of holding postures. Mobility also affects gaze interaction: Lei et al.~\cite{lei2023DynamicRead} reported higher gaze-estimation error when users walked compared with sitting, motivating motion-adaptive techniques for mobile contexts.

% While these studies provide valuable insights, many were conducted in controlled or static environments, which may not fully capture the range of movements and usage patterns associated with handheld devices in real-world settings. Our research aims to address this gap by focusing specifically on dynamic mobile scenarios, exploring how motion impacts gaze estimation accuracy during natural interactions, and utilized Lasson regression process to quantify the effects of motion in each direction.

% Taken together, prior studies establish that head pose, viewing distance, holding posture, and mobility can all reduce gaze-estimation accuracy. What remains less clear is how these factors combine during \emph{semi-controlled handheld interaction}, how they differ between within-task stability and task transitions, and how they interact with calibration-based deployment. Our work addresses this narrower question by analysing a front-camera (RGB) 2D gaze-estimation pipeline across multiple motion conditions and calibration regimes. We use sparse regression as an interpretability-oriented tool to rank motion-related factors, not to claim universal coefficients, but to characterise which factors are most associated with error under the present design.

Taken together, prior studies establish that head pose, viewing distance, holding posture, image quality, and mobility can all affect gaze-estimation accuracy. What remains less clear is how these factors combine during \emph{semi-controlled handheld interaction}, how they differ between sustained task use and task transitions, and how they interact with calibration-based deployment. Our work addresses this narrower question by analysing a front-camera RGB 2D gaze-estimation pipeline across multiple motion conditions and calibration regimes, using regression-based factor analysis as an interpretability tool to identify which geometry- and motion-related signals are most closely associated with error in the present handheld evaluation setting.

\section{Eye Tracking System}
\label{sec:systemdesign}

% Our work aims to understand interactive impact factors between the prediction of 2D gaze estimation and body motion. To support this analysis by empirical evidence, we design and develop a data collection framework for mobile gaze estimation that simultaneously collects eye tracking and motion data. 
To study motion-related degradation in handheld 2D gaze estimation, we required an instrumentation framework that could capture three streams under the same protocol: front-camera observations for gaze estimation, calibration and test events for measuring error, and motion signals describing changes in user-device geometry. The framework is therefore not presented as a new gaze-estimation algorithm; it is the experimental infrastructure that lets us connect behaviour, calibration, and error within a single handheld setup.

\begin{figure}[!htbp]
    \centering
    \includegraphics[width=0.8\textwidth]{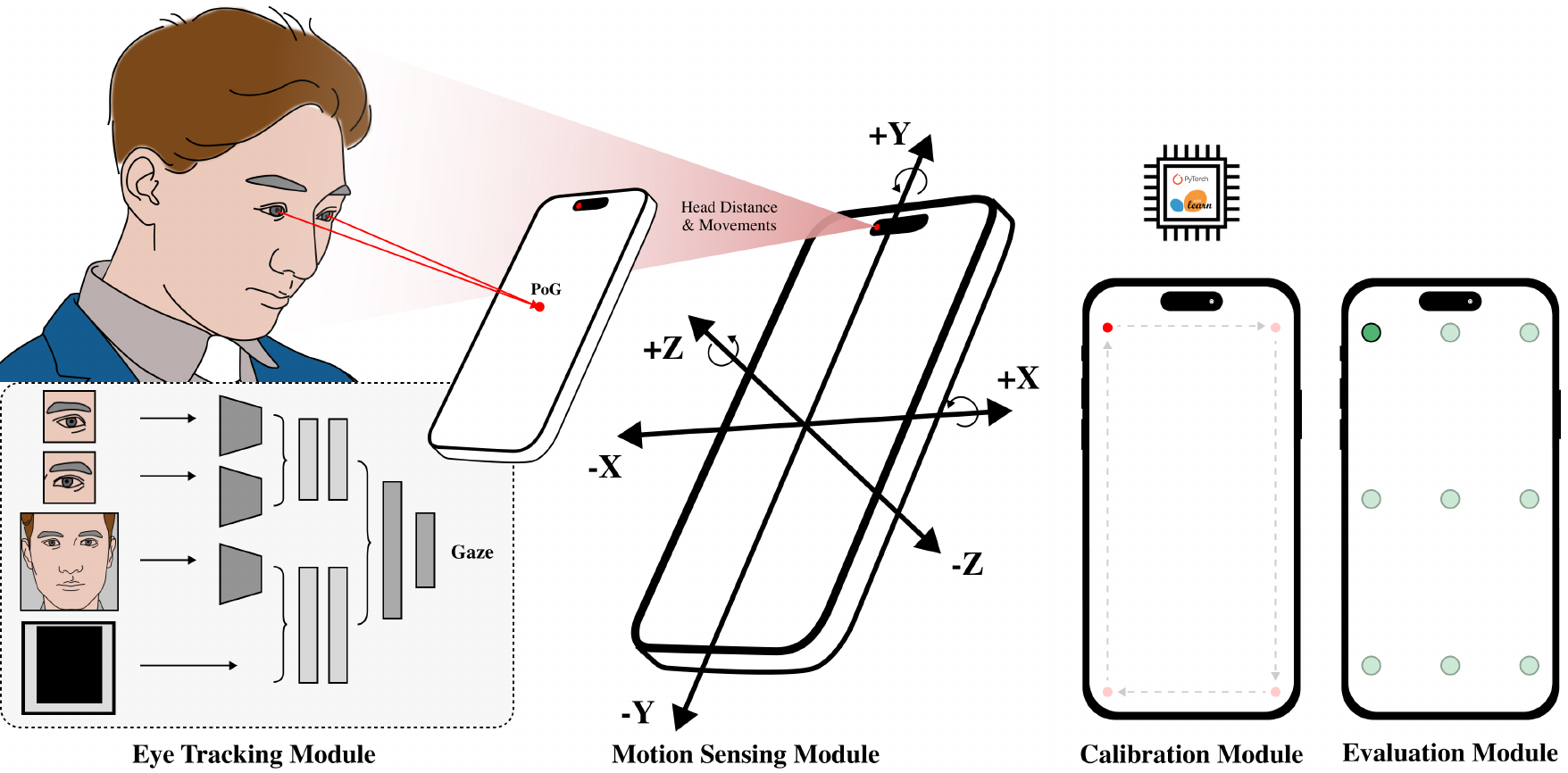}
    \caption{Overview of the experimental mobile eye-tracking system used to synchronise front-camera RGB input, calibration and evaluation events, and motion-sensing data.}
    \label{fig:system-comp}
\end{figure}

Figure~\ref{fig:system-comp} shows the four modules of the framework. The Eye Tracking Module acquires front-facing RGB frames and runs a representative 2D gaze-estimation pipeline. The Calibration Module collects short user-specific calibration sequences and trains lightweight calibrators. The Evaluation Module presents gaze targets and records prediction error. The Motion Sensing Module records IMU signals and camera-derived head measurements, which are later used to analyse geometry change and identify motion-related factors.

\subsection{Eye Tracking Module}\label{subsec:eye_tracking_module}

The Eye Tracking Module provides the visual signal used for 2D gaze estimation. We adopted a client-server architecture, similar to established handheld mobile eye-tracking systems~\cite{lei2023DynamicRead, kong2021eyemu}. The client, implemented in Flutter~\cite{flutter2024}, runs on the mobile device and captures camera frames, sensor streams, and task events. The server runs real-time model inference using PyTorch and scikit-learn~\cite{pytorch2024, sklearn2024}. 
% The client and server communicate over Wi-Fi using Transmission Control Protocol (TCP), allowing the mobile device to remain responsive while computationally heavier inference is performed on the server.}

% Our module interfaces with the device's front-facing camera, capturing video at 30 frames per second (fps). Each frame undergoes processing to detect facial landmarks using Google's ML Kit~\cite{Google2023mlkit}. This step typically takes 10-25ms per frame. As our focus is to study the key impact factors in 2D appearance-based gaze estimation, especially head-to-screen distances, head movement, and device orientation, we intentionally do not rectify head positions in facial images. This decision differs from the common practice in 3D gaze estimation pipeline where head position rectification is often applied to address head movement variability~\cite{zhang18revisiting}.
The module uses the device's \emph{front-facing RGB camera}, capturing video at 30 frames per second (fps). Each frame is processed with Google's ML Kit~\cite{Google2023mlkit} to detect facial landmarks. Because the purpose of this study is to examine how a practical 2D appearance-based pipeline behaves under changing handheld geometry, we retain naturally occurring head-pose and device-orientation variation rather than rectifying facial images into a canonical pose. This differs from 3D normalization pipelines, where head pose is often explicitly estimated and normalised before gaze prediction~\cite{zhang18revisiting}.

The server runs a representative 2D gaze-estimation model based on the iTracker family of architectures~\cite{krafka2016eye}. The model was pre-trained on GazeCapture~\cite{krafka2016eye} and fine-tuned on device-specific RGBDGaze data~\cite{arakawa2022rgbdgaze}. 
% We use this model as a practically relevant front-camera RGB baseline rather than as a state-of-the-art claim. 
Its average Euclidean error on the held-out device-specific test set is 2.05\,cm, which makes it suitable as a baseline predictor and feature extractor for the calibration analyses described below.

\subsection{Calibration Module}\label{subsec:calibration_module}

The Calibration Module collects short user-specific calibration data and trains a lightweight correction model. During calibration, users fixate on a moving dot that traverses the screen boundary. The system synchronises the dot location with the corresponding front-camera frame and model output, producing paired samples of visual features and known on-screen gaze targets. Based on pilot testing and prior mobile gaze work~\cite{kong2021eyemu, lei2023DynamicRead}, we use a 5-second calibration cycle, which typically yields 125--150 calibration pairs while keeping the procedure short enough for repeated use in a handheld study.

For each calibration sequence, the gaze-estimation model first extracts high-dimensional visual features from the first fully connected layer of the network. These features are then paired with the ground-truth target locations and used to train a Support Vector Regression (SVR) calibrator. At test time, the SVR replaces the original final regression layers for participant- and condition-specific gaze prediction. We chose SVR because it is a lightweight and commonly used calibration method in mobile gaze-estimation pipelines~\cite{lei2023DynamicRead, krafka2016eye, valliappan2020accelerating}, with support from comparative calibration studies~\cite{alexiev2019enhancing}.

\subsection{Evaluation Module}\label{subsec:evaluation_module}

The Evaluation Module measures gaze-estimation error after calibration. It displays target dots at predefined screen locations in random order and logs the target location, predicted gaze point, timestamp, camera frame, and synchronised motion signals. Although the interface can vary the number and sequence of target dots, this study uses a 9-point evaluation layout so that prediction error is sampled across different screen regions. These evaluation sequences provide the RMSE values analysed in Section~\ref{sec:results}.

\subsection{Motion Sensing Module}\label{subsec:motion_sensing_module}

The Motion Sensing Module records the geometry and motion signals used to interpret gaze-estimation error. It combines the device's built-in IMU with front-camera landmark measurements. Using Flutter's \texttt{sensor\_plus} package and Google's ML Kit, the module records raw and processed acceleration, device rotation, ambient magnetic-field readings, head pose, and head-to-camera distance. Table~\ref{tab:motionsensormoduleFunction} summarises the collected and estimated signals.

\textbf{Device movement:} The module records device manipulation through IMU-derived orientation and movement signals. These measurements capture changes in how the phone is held, tilted, and moved during task performance, calibration, and testing.

\textbf{Head movement:} The front camera provides head-related measurements from facial landmarks, including head pose and the apparent scale of the eye region. Head pose is represented as Euler angles, specifically pitch and yaw relative to the camera.

\textbf{Head-to-camera distance:} Throughout the paper, ``head distance'' refers to the estimated distance between the user's eye region and the \emph{front camera}. Because the front camera is mounted near the display, this measure is a practical proxy for viewing-distance change, but it is not identical to physical eye-to-screen distance. We therefore use the more precise term \emph{head-to-camera distance} when referring to the value estimated from facial landmarks.

% Since no out-of-the-box API measures head distance, we developed a custom method based on the linear proportional relationship between interpupillary distance and the distance from the eyes to the phone screen. We manually measured the physical interpupillary distance ($IOD_{\text{Physical}}$) and the distance from the interpupillary centre to the camera ($D_{\text{Head-Camera}}$) using a ruler. The interpupillary distance within the image frame ($IOD_{\text{Frame}}$) is then calculated based on facial landmarks detected in the images. These measurements are used to establish a regression model, as shown in Eq.~\ref{eq:headdistance}, to learn the coefficients $\beta_0$, $\beta_1$, and the random error $\epsilon$, allowing us to approximate $D_{\text{Head-Camera}}$. This method is implemented in the motion sensing module to provide real-time estimations of head-to-camera distance, as depicted in Figure~\ref{fig:eye-distance}.
No standard mobile API directly provides head-to-camera distance, so we implemented a participant-specific estimator based on the approximately linear relationship between interpupillary scale in the image and camera distance. We manually measured each participant's physical interpupillary distance ($IOD_{\text{Physical}}$), computed the interpupillary distance observed in each image frame ($IOD_{\text{Frame}}$), and fitted the regression model in Eq.~\ref{eq:headdistance}. The resulting estimate is used as a proxy for handheld viewing-distance change in the behavioural and factor analyses.

\begin{equation}
    D_{\text{Head-Camera}} = \beta_0 + \beta_1 \times \frac{IOD_{\text{Physical}}}{IOD_{\text{Frame}}} + \epsilon
    \label{eq:headdistance}
\end{equation}

\begin{figure}[!htbp]
    \centering
    \includegraphics[width=0.6\textwidth]{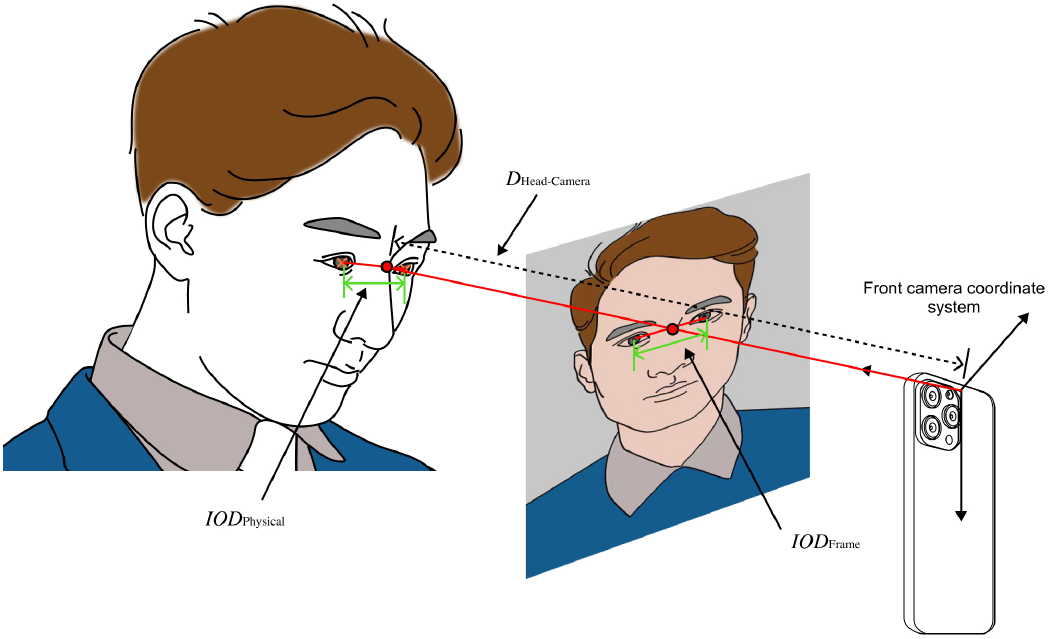}
    \caption{Head-to-camera distance estimation from the relationship between physical interpupillary distance and image-frame interpupillary scale.}
    \label{fig:eye-distance}
\end{figure}

\begin{table}[!htbp]
\centering
\resizebox{\textwidth}{!}{%
\begin{tabular}{cccl}
\hline
\textbf{Sensor Events} & \textbf{Frequency} & \textbf{Dimension} & \multicolumn{1}{c}{\textbf{Description}} \\
\hline
UserAccelerometer & 50Hz & X,Y,Z & Acceleration (m/s\textsuperscript{2}) without the effects of gravity. \\
Accelerometer & 50Hz & X,Y,Z & Acceleration (m/s\textsuperscript{2}) with the effects of gravity. \\
Gyroscope & 50Hz & X,Y,Z & Device rotation. \\
Magnetometer & 50Hz & X,Y,Z & Ambient magnetic field surrounding the device. \\
Head distance & 30Hz & distance & Estimated distance between the eye region and the front camera. \\
Head pose & 30Hz & Pitch, Yaw & Head orientation relative to the camera. \\
\hline
\end{tabular}
}
\caption{Sensor data collected or estimated by the Motion Sensing Module.}
\label{tab:motionsensormoduleFunction}
\end{table}

\section{Study Design}\label{sec:studydesign}
To investigate how motion-related geometry change affects handheld 2D gaze estimation, we organised the study around three sub-questions:

\begin{itemize}
    \item[Q1] - \textbf{Motion Pattern Exploration}: What motion and geometry patterns emerge when users perform everyday-style tasks on handheld mobile devices?
    \item[Q2] - \textbf{Motion Impact on Performance}: How do motion condition, task context, and calibration-test mismatch affect the accuracy of a front-camera RGB 2D gaze-estimation pipeline?
    \item[Q3] - \textbf{Factor Identification}: Which geometry- and motion-related variables are most closely associated with gaze-estimation error under different handheld usage regimes?
\end{itemize}

% To address these questions, we designed the user studies depicted in Figure~\ref{fig:study-outline}. The User Study 1 focuses on understanding motion patterns and behaviours during typical user interaction tasks, while the User Study 2 links these motions to gaze estimation accuracy, providing a foundation for subsequent analysis.

\begin{figure}[!htbp]
    \centering
    \includegraphics[width=1\textwidth]{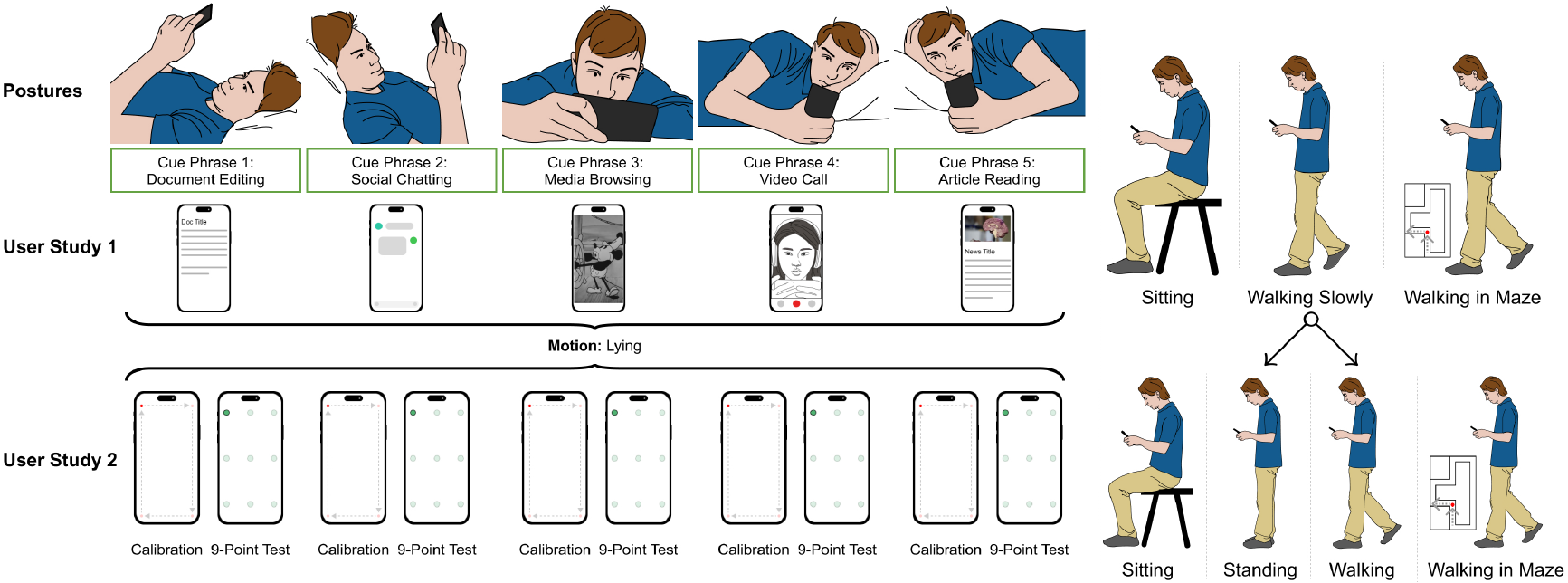}
    \caption{Schematic overview of the user study design. User Study~1 characterises handheld geometry and motion patterns during everyday-style tasks. User Study~2 evaluates 2D gaze-estimation performance under task-elicited semi-controlled handheld conditions.}
    \label{fig:study-outline}
\end{figure}

\paragraph{Design overview}
Figure~\ref{fig:study-outline} summarises the two-study design. We separated behavioural observation from gaze-error evaluation because accurate error measurement requires calibration and target-fixation tasks, which can suppress some of the natural motion variation that we wish to characterise. User Study~1 therefore focuses on motion and posture regularities during everyday-style handheld tasks, while User Study~2 evaluates how comparable task-elicited motion patterns interact with calibration and 9-point gaze testing.

Both studies used a within-subject design with the same 10 participants completing all assigned conditions. In User Study~1, the primary independent variables were \emph{motion condition} (lying, sitting, walking slowly, walking in maze) and \emph{task type} (document editing, social chatting, social media browsing, video calling, news reading). The dependent variables were sensor-derived measures of handheld geometry and motion, including head-to-camera distance, head pose, device orientation, IMU features, and cluster-based motion-pattern structure.

In User Study~2, the primary independent variables were \emph{motion condition} (lying, sitting, standing, walking, walking in maze), \emph{task type}, \emph{calibration regime} (base model, one-off, task-specific, motion-specific), and \emph{test relation} (within-task/between-task; within-motion/between-motion). The dependent variables were gaze-estimation error (RMSE in cm), temporal drift after calibration, and feature-group rankings estimated by regression-based analysis. We chose a within-subject design to reduce inter-participant variability in anthropometrics and phone-use style, at the cost of a relatively long session; sample-size scope is discussed in the Limitations subsection of Section~\ref{sec:discussion}.

\subsection{User Study 1 -- Interaction Pattern}
% The objective of user study 1 is to analyse the spatial relationships and motion behaviours exhibited by users while interacting with their mobile devices during everyday tasks.  This involves observing and recording a variety of motion states such as lying, sitting, standing, walking slowly, and navigating through a maze, to simulate real-world scenarios. These conditions are chosen based on an extensive literature review~\cite{huang2017screenglint,arakawa2022rgbdgaze, zhang2015appearance,huang2017tabletgaze, funes2014eyediap, lei2023DynamicRead, team2024screentime} and are further supplemented by our maze navigation condition to simulate complex user movements encountered in daily life. 

User Study~1 was designed to observe how user-device geometry changes while participants perform everyday-style tasks on their own phones. We recorded four motion conditions: \emph{lying}, \emph{sitting}, \emph{walking slowly}, and \emph{walking in a maze}. A separate standing-only block was not included in this observational study because participants used their own phone with the experimental device attached behind it, and adding another prolonged posture block would have increased the burden of an already long session. Standing was instead introduced in User Study~2, where participants used the experimental device directly, allowing us to separate non-locomotive upright posture from walking during gaze-evaluation procedures.

% To ensure ecological validity, we selected tasks that mimic daily usage patterns on handheld devices. Specifically, we asked participants to perform a series of routine tasks under each condition including, \emph{document editing} simulates note-taking or drafting emails, \emph{social chatting} represents quick message replies via popular apps like WhatsApp, \emph{social media browsing} (e.g., Instagram, Twitter) captures casual content consumption, \emph{video calling} allows for real-time communication or short call simulations, and \emph{news reading} aligns with web-based article consumption, with cue phrases provided to guide task execution as shown in Table~\ref{tab:p1tasks}. These tasks were chosen for their prevalence in everyday life and for covering a spectrum of interactions (e.g., heavy typing, continuous scrolling, periodic glances). 

To approximate common mobile behaviours while remaining feasible in a laboratory setting, we selected five cue-guided tasks: document editing, social chatting, social media browsing, video calling, and news reading (Table~\ref{tab:p1tasks}). These tasks were chosen to elicit different interaction styles, including typing, scrolling, visual scanning, and conversational attention shifts.

To reduce order effects, tasks were randomised for each participant. Participants received a short 1--2 minute familiarisation session with the experiment application before starting, and were encouraged to use their natural holding posture, such as one-handed or two-handed use.

\begin{table}[!htbp]
\centering
\resizebox{\textwidth}{!}{%
\begin{tabular}{l|l} 
\hline
Task & Cue Phrase \\
\hline
Document editing & \begin{tabular}[c]{@{}l@{}}``Create a new document using your preferred document editing \\app and write today's diary entry.''\end{tabular} \\ 
\hline
Social chatting & \begin{tabular}[c]{@{}l@{}}``Open your preferred social media app, such as WhatsApp, and respond\\~to any unread messages.''\end{tabular} \\ 
\hline
Social media browsing & ``Open a social media app like Instagram to browse content or like posts.'' \\ 
\hline
Video calling & \begin{tabular}[c]{@{}l@{}}``Use your preferred video calling app to either simulate a video call or \\engage in a brief call with a friend.''\end{tabular} \\ 
\hline
News reading & ``Open your preferred news app to read some articles.'' \\
\hline
\end{tabular}
}
\caption{Cue phrases for everyday-style task prompts.}\label{tab:p1tasks}
\end{table}

Data collection used the Motion Sensing Module described in Section~\ref{subsec:motion_sensing_module}, capturing accelerometer, gyroscope, magnetometer, device-orientation, head-pose, and head-to-camera distance measurements for the behavioural analyses in Section~\ref{sec:results}.

\subsection{User Study 2 -- Impact and Factors}\label{subsec:impact_factor}
% User Study 2 aims to assess the performance of a 2D gaze estimation method under simulated real-world conditions by correlating the frequency and intensity of motion patterns with model prediction errors. This study extends the first by examining how well the current mobile eye-tracking systems perform under typical everyday tasks (see Table~\ref{tab:p1tasks}). We replicate cue-guided tasks from user study 1 to evoke similar postures and behaviours under controlled motion conditions, including lying, sitting, standing, walking, and walking in maze. Each participant produces 25 pairs of calibration and test data, corresponding to 5 motion conditions and 5 cue-guided tasks. For each cue-guided task of a motion condition, participants were subjected to a calibration and 9-point test procedure. This collected dataset enables a systematic evaluation of how factors such as head-to-screen distance, head movements, and device orientation affect gaze estimation accuracy in dynamic, mobile contexts.

User Study~2 evaluated the same front-camera RGB 2D gaze-estimation pipeline under \emph{semi-controlled, task-elicited} handheld conditions. Participants completed cue-guided tasks similar to those in User Study~1, but also performed explicit calibration and 9-point gaze testing. The study included five motion conditions: lying, sitting, standing, walking, and walking in a maze. Each participant produced 25 calibration-test pairs, corresponding to 5 motion conditions and 5 task types. This dataset supports the analysis of how head-to-camera distance, head movement, device-orientation change, calibration regime, and calibration-test mismatch relate to gaze-estimation error.

\subsection{Comparison of User Study 1 and User Study 2}
The two studies play complementary roles. User Study~1 provides a behavioural reference for how handheld geometry changes during everyday-style tasks when participants use their own phones. User Study~2 provides the gaze-error measurements needed to evaluate the 2D gaze-estimation pipeline, but the calibration and 9-point testing procedures make it more constrained than the observational setting. In particular, participants may hold the device more steadily during calibration and target fixation than they would during unconstrained use. We therefore interpret User Study~1 as a reference for broader task-related geometry variation, and User Study~2 as a constrained deployment study for examining how those motion factors interact with calibration and gaze-estimation error.

% The inclusion of calibration and testing in User Study 2 means that participants may adjust their interaction patterns slightly due to the demands of the gaze estimation tasks. However, by replicating the tasks and conditions from User Study 1, we aim to maintain consistency in the behaviours elicited.

% This dual-study design allows us to first characterize the motion patterns in natural usage (addressing Q1) and then directly assess their impact on gaze estimation performance (addressing Q2 and Q3). By comparing the findings from both studies, we can identify which aspects of user motion and behaviour are most critical for gaze estimation accuracy and develop strategies to mitigate their negative effects.

\section{User Study and Data Collection}\label{sec:userstudy}

This section reports the participant sample, apparatus, experimental environment, study procedures, and preprocessing steps used to produce the behavioural and gaze-evaluation datasets analysed in Section~\ref{sec:results}. The research protocol was approved by the university ethics committee (reference number CS16900).

\subsection{Participants}

% A cohort of 10 participants (4 females, 6 males) was recruited from the undergraduate and postgraduate communities at our university. The sample's age distribution ranged from 21 to 31 years (M = 26.4, SD = 3.14). All participants possessed either normal or corrected-to-normal vision. Among the participants, 50\% (n = 5) wore corrective lenses, with mean myopia measurements of 299 diopters (SD = 96.45) for the left eye and 334 diopters (SD = 148.40) for the right eye.

% Given the study's focus on spatial and motion impacts, we collected additional anthropometric data. Participants' heights ranged from 155 cm to 185 cm (M = 172.5 cm, SD = 10.55). The arm length of the dominant hand varied from 52 cm to 63 cm (M = 57.2 cm, SD = 3.96), while interpupillary distances ranged from 6.02 cm to 6.50 cm (M = 6.32 cm, SD = 0.23).
% A demographic questionnaire assessed participants' technology familiarity and usage patterns. On a 5-point Likert scale (1 = least familiar, 5 = most familiar), participants reported high familiarity with mobile phone usage (M = 4.5, SD = 0.81) but low familiarity with gaze technologies (M = 1.5, SD = 0.5). The frequency of mobile phone use while walking was moderate (M = 2.9, SD = 1.04, on a scale where 1 = lowest frequency and 5 = highest frequency).

A cohort of 10 participants (4 female, 6 male) was recruited from the undergraduate and postgraduate communities at our university, and the \emph{same 10 participants completed both User Study~1 and User Study~2}. Participants were between 21 and 31 years old (\(M = 26.4\), \(SD = 3.14\)) and all had normal or corrected-to-normal vision. Participants received \pounds40 compensation for completing the full session, which combined both studies and typically lasted slightly over two hours including briefing, setup, calibration, task execution, and breaks.

Given the study's focus on spatial and motion impacts, we collected additional anthropometric data. Participants' heights ranged from 155\,cm to 185\,cm (\(M = 172.5\), \(SD = 10.55\)), dominant-arm length ranged from 52\,cm to 63\,cm (\(M = 57.2\), \(SD = 3.96\)), and interpupillary distance ranged from 6.02\,cm to 6.50\,cm (\(M = 6.32\), \(SD = 0.23\)). Among the participants, 50\% (\(n=5\)) wore corrective lenses.

A demographic questionnaire assessed participants' technology familiarity and usage patterns. On a 5-point Likert scale (1 = least familiar, 5 = most familiar), participants reported high familiarity with mobile phone usage (\(M = 4.5\), \(SD = 0.81\)) but low familiarity with gaze technologies (\(M = 1.5\), \(SD = 0.5\)). The frequency of mobile phone use while walking was moderate (\(M = 2.9\), \(SD = 1.04\)).

The scope of this participant sample is discussed in the Limitations subsection of Section~\ref{sec:discussion}.

\subsection{Apparatus and Environmental Setting}

The experimental setup used the client-server system described in Section~\ref{sec:systemdesign}. The client device was an Apple iPhone 13 Pro Max (6.7-inch display, 240 grams, 256 GB storage, iOS 16). It was paired with a server equipped with an Intel i9-13900HX processor, NVIDIA RTX 4080 GPU, and 32GB RAM for real-time processing and data storage. The system achieved a gaze-inference frequency of 25--30 fps.

Both user studies were conducted in a large lecture room measuring 8\,m $\times$ 20\,m. A 5\,m $\times$ 10\,m rectangular area was arranged as a concave walking maze, with the walking path enclosed by soft cushions. Chairs and loungers were placed in a 3\,m $\times$ 3\,m area near the maze so that static and walking conditions could be completed in the same room. The room was lit by natural light supplemented with artificial illumination.

\subsection{Experiment Procedure of User Study 1: Understanding User-Device Interaction Patterns}

Before starting, participants were briefed on the purpose, apparatus, and procedure of both studies.
% To ensure a natural interaction and mitigate privacy concerns, participants were instructed to use their personal phones for the tasks prompted by cue phrases. Simultaneously, our experimental phone was affixed to the rear of their device to capture relevant data, including IMU and facial images from the camera. However, we recognise a potential drawback of this approach: the combined weight of the two phones might alter participants' handling and mobility of their devices.
To approximate participants' familiar interaction habits while mitigating privacy concerns, participants were instructed to use their own phones for the prompted tasks, while our experimental phone was affixed to the rear of their device to capture IMU signals and front-camera observations. This design preserved participants' preferred applications and content, but it also introduced an important trade-off: the combined weight and thickness of the two devices may have altered handling behaviour. We therefore treat User Study~1 as an \emph{everyday-style but still instrumented} observational study rather than as fully unconstrained natural use.

Participants completed the five cue-guided task types introduced in Section~\ref{sec:studydesign}: document editing, social chatting, social media browsing, video calling, and news reading. The cue phrases guided the activity while still allowing participants to use familiar applications and interaction styles. For example, participants could use their preferred social media or document-editing application, and headphones could be worn during chat or video-call tasks to protect privacy.

The Study~1 procedure is summarised in Figure~\ref{fig:study-outline}. Participants first entered the lying condition, where they lay on a soft sofa in the experiment room. Once the participant was positioned, the system began recording. The participant then completed the cue-guided tasks under each motion condition. Each task lasted approximately 4--6 minutes, and each motion-condition block lasted approximately 15--20 minutes. User Study~1 took approximately 60--90 minutes per participant.
After cleaning, User Study~1 yielded 1,232,633 IMU events for each of the accelerometer, gyroscope, and magnetometer streams, and 739,578 camera-derived head-movement and head-pose samples across 40 trials (10 participants \(\times\) 4 motion conditions), as summarised in Table~\ref{tab:exp1-sensor-data}.

\begin{table}[H]
\centering
{\footnotesize
\begin{tabular}{c|ccccc}
\hline
\diagbox{Sensor}{Motion} & Lying & Sitting & Walking Slowly & Walking in Maze & Total \\
\hline
IMU & 311,438 & 310,802 & 324,690 & 285,703 & 1,232,633 \\
Camera & 186,862 & 186,481 & 194,814 & 171,421 & 739,578 \\
\hline
\end{tabular}}
\caption{Distribution of data samples across motion conditions and sensor types in User Study~1.}
\label{tab:exp1-sensor-data}
\end{table}

\subsection{Experiment Procedure of User Study 2: Impact and Factor Analysis}

User Study~2 evaluated the front-camera RGB 2D gaze-estimation pipeline under semi-controlled, task-elicited handheld conditions. Unlike User Study~1, participants used the experimental device directly so that the calibration and 9-point testing interface could be controlled precisely. For each task and motion condition, participants completed a calibration sequence followed by a 9-point gaze test while holding and interacting with the device under the prompted posture or motion condition.

Across 50 trials (10 participants \(\times\) 5 motion conditions), we collected 803,241 time-stamped events. Table~\ref{tab:exp3-traintest-sensor-data} reports the number of IMU and camera samples by motion condition and experimental phase. In total, the dataset contains 137,011 IMU and 82,205 camera events during calibration, and 666,230 IMU and 399,736 camera events during the 9-point gaze tests.

\begin{table}[H]
\centering
{\footnotesize
\begin{tabular}{cc|cccccc}
\hline
\multicolumn{2}{l|}{\diagbox{Phase}{Samples}{Motion}} & Lying & Sitting & Standing & Walking & Walking in Maze & Total \\
\hline
\multirow{2}{*}{Calibration} & IMU  & 27,471 & 27,230 & 27,949 & 27,652 & 26,709 & 137,011 \\
 & Camera  & 16,482 & 16,338 & 16,769 & 16,591 & 16,025 & 82,205 \\
\multirow{2}{*}{9-Point Test} & IMU  & 130,419 & 133,647 & 135,409 & 133,512 & 133,243 & 666,230 \\
 & Camera  & 78,251 & 80,188 & 81,245 & 80,107 & 79,945 & 399,736 \\
\hline
\end{tabular}}
\caption{Distribution of data samples across motion conditions, sensor types, and experimental phases in User Study~2.}
\label{tab:exp3-traintest-sensor-data}
\end{table}

\subsection{Data Preprocessing and Feature Extraction}
The analyses in Section~\ref{sec:results} rely on synchronised camera and motion records. We therefore preprocessed the raw streams to align their timestamps, remove implausible observations, and extract fixed-length feature summaries from the resulting time-series data.

\subsubsection{Preprocessing}\label{subsec:preprocessing}

During preprocessing, IMU sensor events were synchronised with camera frames by timestamp. Because the IMU and camera operate at different sampling frequencies, we used neighbour interpolation~\cite{lepot2017interpolation} to estimate the IMU values corresponding to each camera frame. This produced paired camera and motion observations for downstream behavioural analysis and gaze-error modelling.

We then applied two denoising steps. First, we used a Butterworth low-pass filter to smooth the IMU sensor streams~\cite{anguita2013public}. Second, we filtered unrealistic head-to-camera distance estimates using an ergonomic threshold. Based on the observed distance distribution, a lower one-sided 95\% confidence interval indicated that values above approximately 100\,cm were unlikely to correspond to active screen viewing; such samples were removed from the analysis.

\subsection{IMU Feature Extraction}\label{subsec:feature}
To summarise the high-dimensional time-series data, we adopted a feature-engineering approach commonly used in human activity recognition~\cite{anguita2013public}. The extracted features fall into three broad families: frequency-domain features, time-domain features, and cross-channel features. Table~\ref{tab:feature-summary} summarises the raw signal groups, feature families, and representative examples.

\begin{table*}[!htbp]
\centering
\small
\begin{tabularx}{\textwidth}{l|c|l|X} % The 'X' column automatically wraps text
\hline
\textbf{Signal Group} & \textbf{Raw Channels} & \textbf{Feature Families} & \textbf{Example Features} \\
\hline
User accelerometer & X, Y, Z & Time, Frequency, Cross-channel & mean, std, range, zero-crossing, FFT mean, top-4 amplitudes, cross-correlation \\
Accelerometer & X, Y, Z & Time, Frequency, Cross-channel & mean, variance, kurtosis, DC component, cosine similarity, synthetic mean \\
Gyroscope & X, Y, Z & Time, Frequency, Cross-channel & mean, std, max/min, FFT peaks, lagged cross-correlation \\
Magnetometer & X, Y, Z & Time, Frequency, Cross-channel & mean, variance, skewness, DC component, cosine similarity \\
Head pose & Pitch, Yaw & Time, Cross-channel & mean, std, range, pitch-yaw correlation \\
Device orientation & Pitch, Yaw, Roll & Time, Cross-channel & mean, std, range, synthetic mean, pairwise correlation \\
Head-to-camera distance & Distance & Time & mean, std, min/max, range, slope \\
Derived motion summaries & -- & Composite & hold angle, velocity, synthetic mean \\
\hline
\end{tabularx}
% \caption{Summary of extracted features used in the motion analysis and regression-based factor identification. The full feature set contains 552 dimensions aggregated across raw channels and feature families.}
\caption{Summary of extracted features used in the motion analysis and regression-based factor identification. The full feature set contains 552 dimensions aggregated across raw channels and feature families.}

\label{tab:feature-summary}
\end{table*}

\textbf{Frequency Domain Features} describe periodic structure in the sensor streams. These features are derived through the transformation of time-domain data into the frequency spectrum using Fast Fourier Transform (FFT). Significant frequency domain features include mean (calculated as $\frac{1}{N} \sum_{i=1}^{N} x_i$, where $x_i$ represents the amplitude of the $i^{th}$ frequency component, and $N$ is the number of frequency components), variance, standard deviation, skewness, kurtosis, direct component (DC), and the top 4 amplitudes along with their corresponding frequencies. Such features are particularly effective in recognising cyclical activities like walking or running.

\textbf{Time Domain Features} describe the distribution and variability of each signal over a time window. They include mean (given by $\frac{1}{N} \sum_{i=1}^{N} x_i$, where $x_i$ is the $i^{th}$ time-domain sample and $N$ is the total number of samples in a window), variance, standard deviation, mode, min/max, range, and zero-crossing rate. These features capture the basic motion attributes including range, intensity, and variability, which are vital for detecting general motion patterns and activity intensity.

\textbf{Cross-channel Features} describe relationships between axes or related signals. For multi-axis (X, Y, Z axes) sensors, it is crucial to analyse the correlations and interactions among these channels. Features in this category include cross-correlation coefficients, represented by $CC_{xy}(k) = \frac{\sum_{i} (a_{x,i} - \bar{a}_x)(a_{y,i+k} - \bar{a}_y)}{\sqrt{\sum_{i}(a_{x,i} - \bar{a}_x)^2 \sum_{i}(a_{y,i+k} - \bar{a}_y)^2}}$, where $a_{x,i}$ and  $a_{y, i+k}$ represents x- and y-axis sensor data at $i$th and $i+k$th (the lag) timestamps, and $\bar{a}_x$ and $\bar{a}_y$ represents the mean x- and y-axis sensor data. This cross-correlation coefficients quantify the degree of similarity and the lead-lag relationship between the time-series data of different axes. Cosine similarity $\frac{\vec{a}_i \cdot \vec{a}_j}{\|\vec{a}_i\|\|\vec{a}_j\|}$ assesses the directional alignment between two multi-axial data vectors $\vec{a}_i$ and $\vec{a}_j$ at the $i$th and $j$th timestamps, providing insight into their relative orientation. The synthetic mean is a composite metric encapsulating the overall magnitude of motion by combining the squared sum of multi-axial signals: $a_i = \sqrt{(a_{x,i})^2 + (a_{y,i})^2 + (a_{z,i})^2}$. These features are instrumental in understanding complex movements, such as the coordination in limb movements during running or multi-axial activities like climbing stairs.

\section{Data Analysis and Results}\label{sec:results}
This section will illustrate the procedure and methodology of data analysis and discuss the main findings on motion patterns. 

\subsection{User Study 1 -- Interaction Pattern}
User Study~1 characterises how users physically interact with a handheld device while performing everyday-style tasks under different motion conditions. The goal of this first analysis is descriptive: before evaluating gaze-estimation error, we first establish what kinds of head, device, and posture variation are present in the behavioural data. We perform two types of analysis: (1) \textit{motion data statistics} where we extract the range and distribution of raw motion signals across motion conditions and (2) \textit{mobility pattern} where we examine whether repeated posture and task-transition patterns can be observed within those signals.

% \begin{itemize}
%     \item \textit{Motion Data Visualisation} examines the range and distribution of raw motion signals across motion conditions.
%     \item \textit{Mobility Patterns} examines whether repeated posture and task-transition patterns can be observed within those signals.
% \end{itemize}

\begin{figure}[!htbp]
    \centering
    \includegraphics[width=0.80\textwidth]{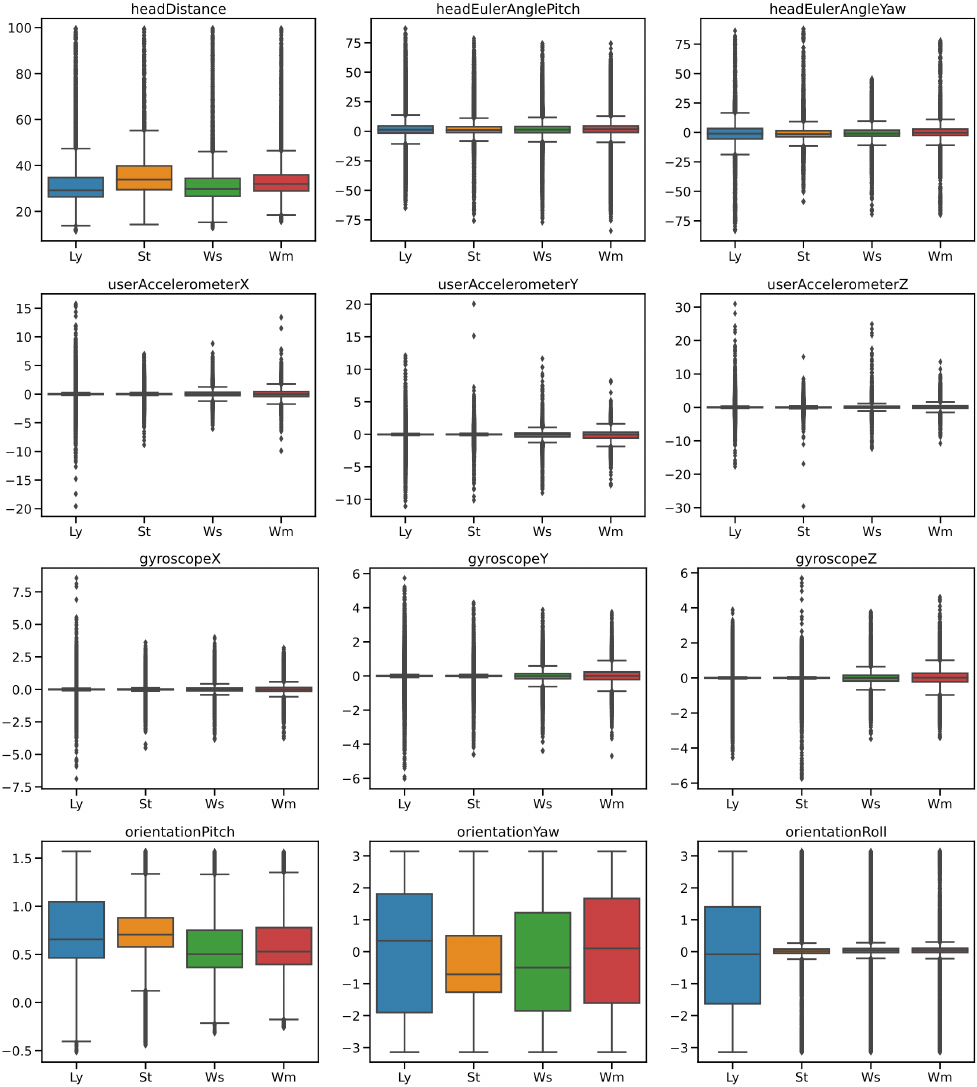}
    \caption{Box plots of sensor data under each motion condition: \textit{Ly} for lying, \textit{St}  for sitting, \textit{Ws} for walking slowly and \textit{Wm} for walking in maze. }
    \label{fig:overall-US1-sensor-data}
\end{figure}

\subsubsection{Motion Data Visualisation}\label{subsec:visualisemotiondata}
\begin{figure}[!htbp]
\centering
\includegraphics[width=0.85\textwidth]{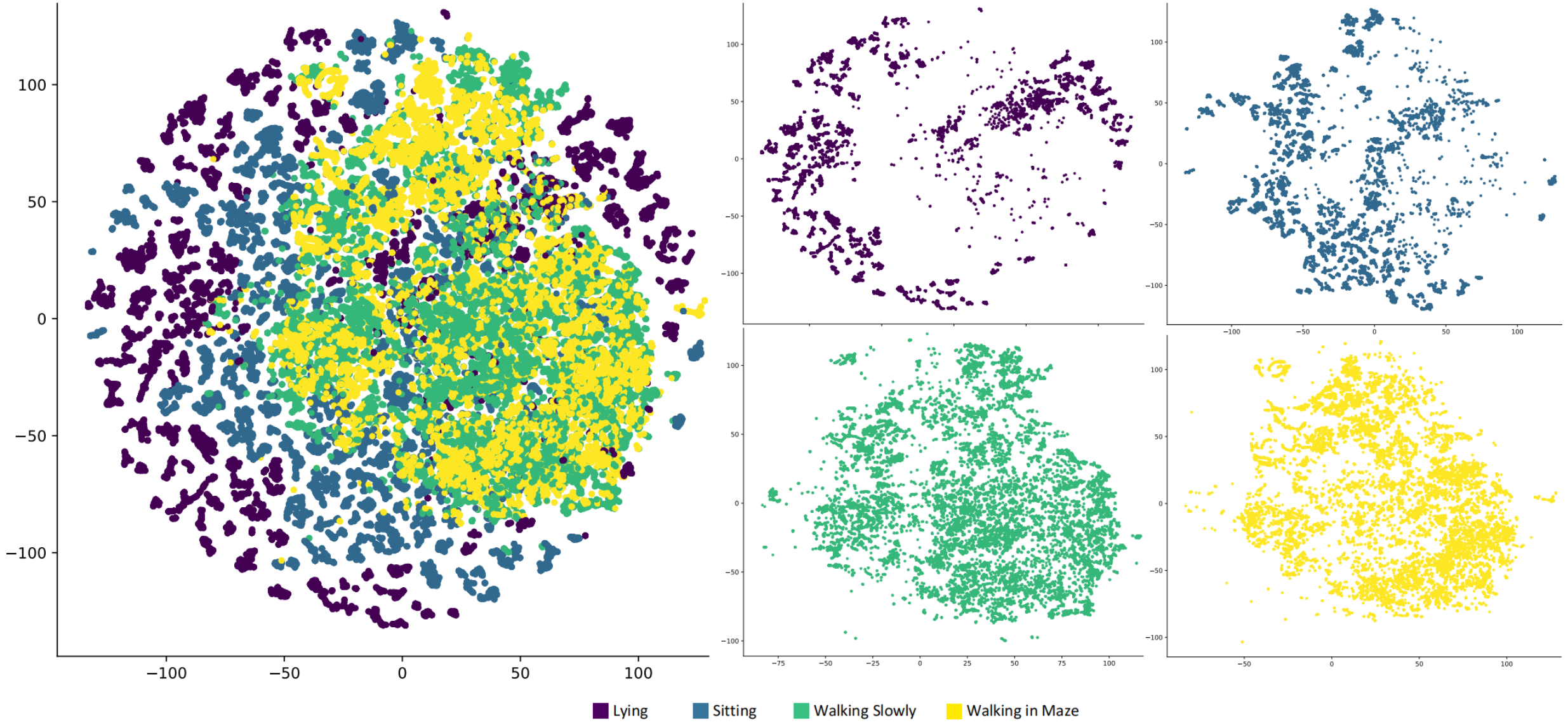}
\caption{t-SNE plots of sensor features under each motion condition}
\label{fig:us1tsneallposture}
\end{figure}

Figure~\ref{fig:overall-US1-sensor-data} summarises the raw motion signals under each motion condition, and Figure~\ref{fig:us1tsneallposture} visualises the corresponding extracted feature distributions. We use these two views to describe both signal-level variation and higher-level separation between motion conditions.

\noindent\textbf{Head distance:} Head distance refers to the estimated head-to-camera distance described in Section~\ref{subsec:motion_sensing_module}. Across User Study~1, the largest variation appeared in lying ($M=31.37$\,cm, $SD=10.09$\,cm), followed by sitting ($M=34.61$\,cm, $SD=8.44$\,cm), walking in maze ($M=33.13$\,cm, $SD=7.57$\,cm), and walking slowly ($M=31.07$\,cm, $SD=7.01$\,cm). This pattern suggests that static postures do not necessarily produce static viewing geometry: when participants are supported by a sofa or chair, they may have more freedom to move their arms, adjust posture, and vary the phone-to-face distance.

\noindent\textbf{Head Euler Angles (Pitch, Yaw):}  Pitch and yaw describe head orientation relative to the camera. Mean pitch values were positive across all conditions, consistent with participants typically holding the phone below eye level, while mean yaw values remained close to 0. More specifically, lying ($M_{pitch} = 1.23$, $SD_{pitch} = 9.61$, $M_{yaw} = -1.24$, $SD_{yaw} = 10.96$), sitting ($M_{pitch} = 2.15$, $SD_{pitch} = 8.09$, $M_{yaw} = -0.73$, $SD_{yaw} = 7.49$), walking slowly ($M_{pitch} = 1.25$, $SD_{pitch} = 7.68$, $M_{yaw} = -0.57$, $SD_{yaw} = 5.20$), and walking in maze ($M_{pitch} = 2.19$, $SD_{pitch} = 10.72$, $M_{yaw} = 0.35$, $SD_{yaw} = 7.34$). 
Variation was again substantial in the two static conditions and in maze walking. That is, when users are finding and avoiding obstacles in the maze, they tend to move their phones out of their viewing range, which is adjust their head-device geometry.

\noindent\textbf{User Accelerometer (X, Y, Z):}  As users are required to interact with the device by various tasks, they are less likely to perform high-intensity activities; therefore, their accelerometer readings remain stable and close to 0: lying ($M_{X} = 0.00$, $SD_{X} =0.49 $, $M_{Y} = -0.05$, $SD_{Y} = 0.38$, $M_{Z} = 0.01$, $SD_{Z} = 0.56$), sitting ($M_{X} = 0.01$, $SD_{X} = 0.32$, $M_{Y} = -0.03$, $SD_{Y} = 0.26$, $M_{Z} = -0.01$, $SD_{Z} = 0.40$), walking slowly ($M_{X} = 0.02$, $SD_{X} = 0.59$, $M_{Y} = -0.10$, $SD_{Y} = 0.55$, $M_{Z} = 0.04$, $SD_{Z} = 0.61$), and walking in maze ($M_{X} = 0.01$, $SD_{X} = 0.72$, $M_{Y} = -0.14$, $SD_{Y} = 0.67$, $M_{Z} = 0.07$, $SD_{Z} = 0.77$). The standard deviations were larger in the two walking conditions than in lying or sitting, indicating that locomotion primarily appears in the accelerometer stream as increased movement variability rather than a shifted mean.

\noindent\textbf{Gyroscope (X, Y, Z):} The gyroscope can obtain data on how fast the orientation of the device changes, and the greater the absolute value of the reading, the faster the angle changes. The mean value of the gyroscope is close to 0 in the three axes for all the motion conditions. but the two walking conditions produced larger standard deviations, especially on the Y and Z axes. This indicates more frequent device-orientation changes during locomotion, with maze walking producing the largest rotational variability. In detail, lying ($M_{X} = 0.00$, $SD_{X} = 0.25$, $M_{Y} = 0.00$, $SD_{Y} = 0.26$, $M_{Z} = 0.00$, $SD_{Z} = 0.21$), sitting ($M_{X} = 0.00$, $SD_{X} = 0.18$, $M_{Y} = 0.00$, $SD_{Y} = 0.18$, $M_{Z} = 0.00$, $SD_{Z} = 0.17$), walking slowly ($M_{X} = 0.00$, $SD_{X} = 0.23$, $M_{Y} = -0.01$, $SD_{Y} = 0.33$, $M_{Z} = -0.01$, $SD_{Z} = 0.39$), and walking in maze ($M_{X} = 0.00$, $SD_{X} = 0.28$, $M_{Y} = 0.01$, $SD_{Y} = 0.40$, $M_{Z} = 0.02$, $SD_{Z} = 0.49$).

\noindent\textbf{Orientation (Pitch, Yaw, Roll):}  The pitch, yaw and roll angles are key indicators which are similar with the head pose to show device orientation. The variation of pitch is similar to that of yaw, and lying has the largest variation, followed by the two walking conditions; more specifically, lying ($M_{Pitch} = 0.74$, $SD_{Pitch} = 0.39$, $M_{Yaw} = 0.02$, $SD_{Yaw} = 1.97$), sitting ($M_{Pitch} = 0.73$, $SD_{Pitch} = 0.29$, $M_{Yaw} = -0.28$, $SD_{Yaw} = 1.47$), walking slowly ($M_{Pitch} = 0.58$, $SD_{Pitch} = 0.30$, $M_{Yaw} = -0.29$, $SD_{Yaw} = 1.80$), and walking in maze ($M_{Pitch} = 0.59$, $SD_{Pitch} = 0.29$, $M_{Yaw} = 0.03$, $SD_{Yaw} = 1.87$). Lying has resulted in a big difference in the mean of Roll and a larger variation than the other conditions; that is, lying ($M_{Roll} =  -0.15$, $SD_{Roll} = 1.68$), sitting ($M_{Roll} = 0.04$, $SD_{Roll} = 0.51$), walking slowly ($M_{Roll} = 0.05$, $SD_{Roll} = 0.28$), and walking in maze ($M_{Roll} = 0.03$, $SD_{Roll} = 0.28$). Device orientation showed a complementary pattern. Lying produced the largest roll variation, while walking conditions produced greater accelerometer and gyroscope variability. In other words, static conditions can still involve large changes in viewing geometry, whereas dynamic conditions more clearly increase movement and rotation signals.

\noindent\textbf{Relationships of Motion Conditions} Figure~\ref{fig:us1tsneallposture} provides a feature-level view of these patterns. Lying and sitting form more distinguishable regions, while walking slowly and walking in maze overlap more strongly, consistent with their shared locomotion signature. The broader spread of the lying condition also aligns with the raw-signal analysis: supported static postures can permit considerable variation in how the device is held relative to the face.

\subsubsection{Mobility Patterns}\label{subsec:mobility-patterns}
\begin{figure}[!htbp]
     \centering
     % \begin{subfigure}[b]{0.45\textwidth}
         \centering
         \includegraphics[width=0.6\textwidth]{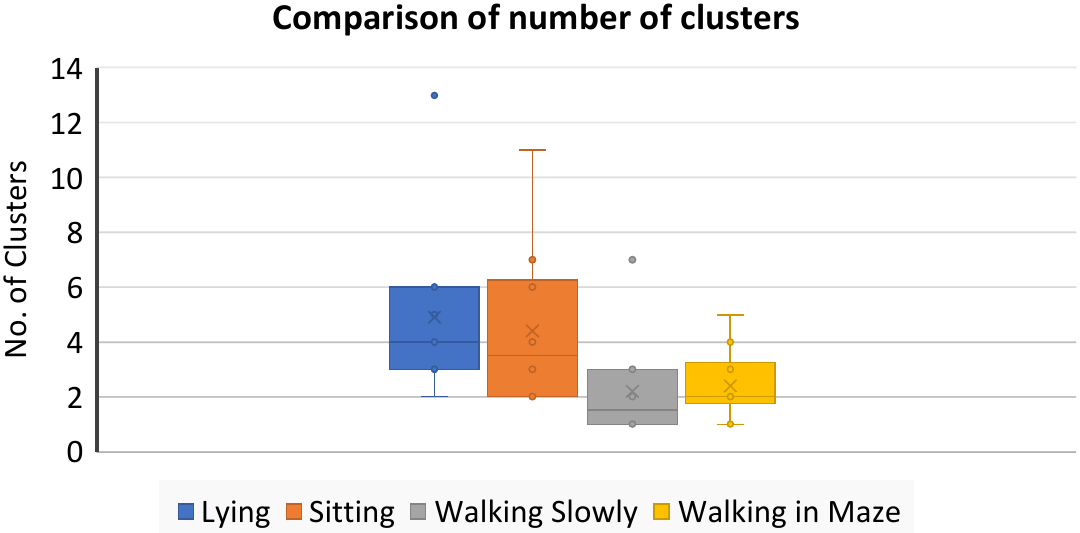}
    \caption{Comparison of the number of motion clusters under different motion conditions}
    \label{fig:gmmhistogram}
\end{figure}

We next examine whether the motion streams contain recurring behavioural patterns rather than only continuous noise. For each participant and motion condition, we fitted Gaussian Mixture Models (GMMs) to the extracted features and counted the resulting clusters. Figure~\ref{fig:gmmhistogram} shows that most participants exhibited $\sim$4 clusters in lying and sitting, but $\sim$2 clusters in the walking conditions. This suggests that static conditions allowed a wider repertoire of postural configurations, whereas walking constrained participants toward fewer, more stable handheld patterns.

\begin{figure}[!htbp]
    \centering
    \includegraphics[width=0.8\textwidth]{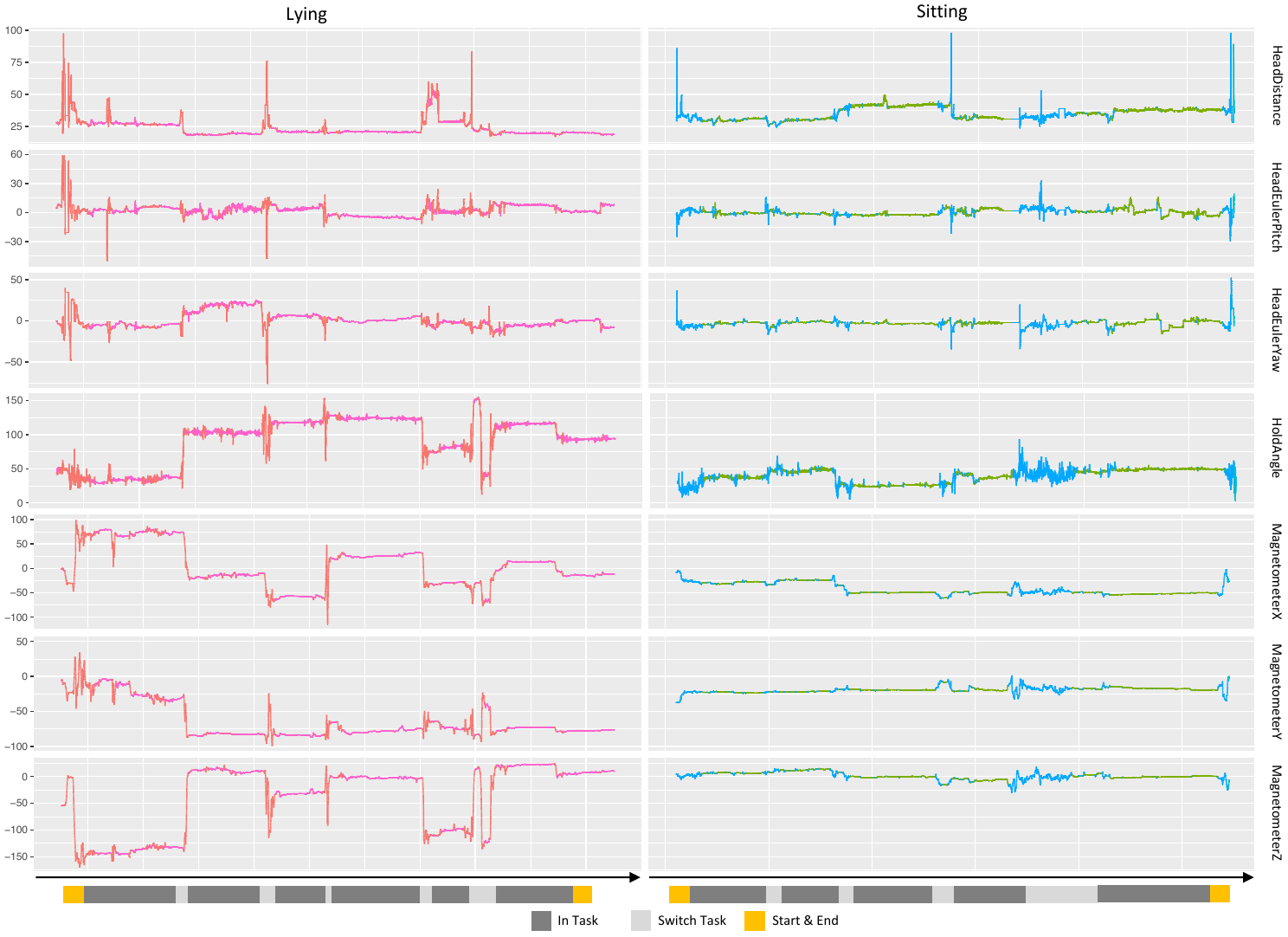}
    \caption{GMM clustering mapped to time-series sensor data on Participant 1 under static conditions}
    \label{fig:exp1gmmtimep1static}
\end{figure}

\begin{figure}[!htbp]
    \centering
    \includegraphics[width=0.8\textwidth]{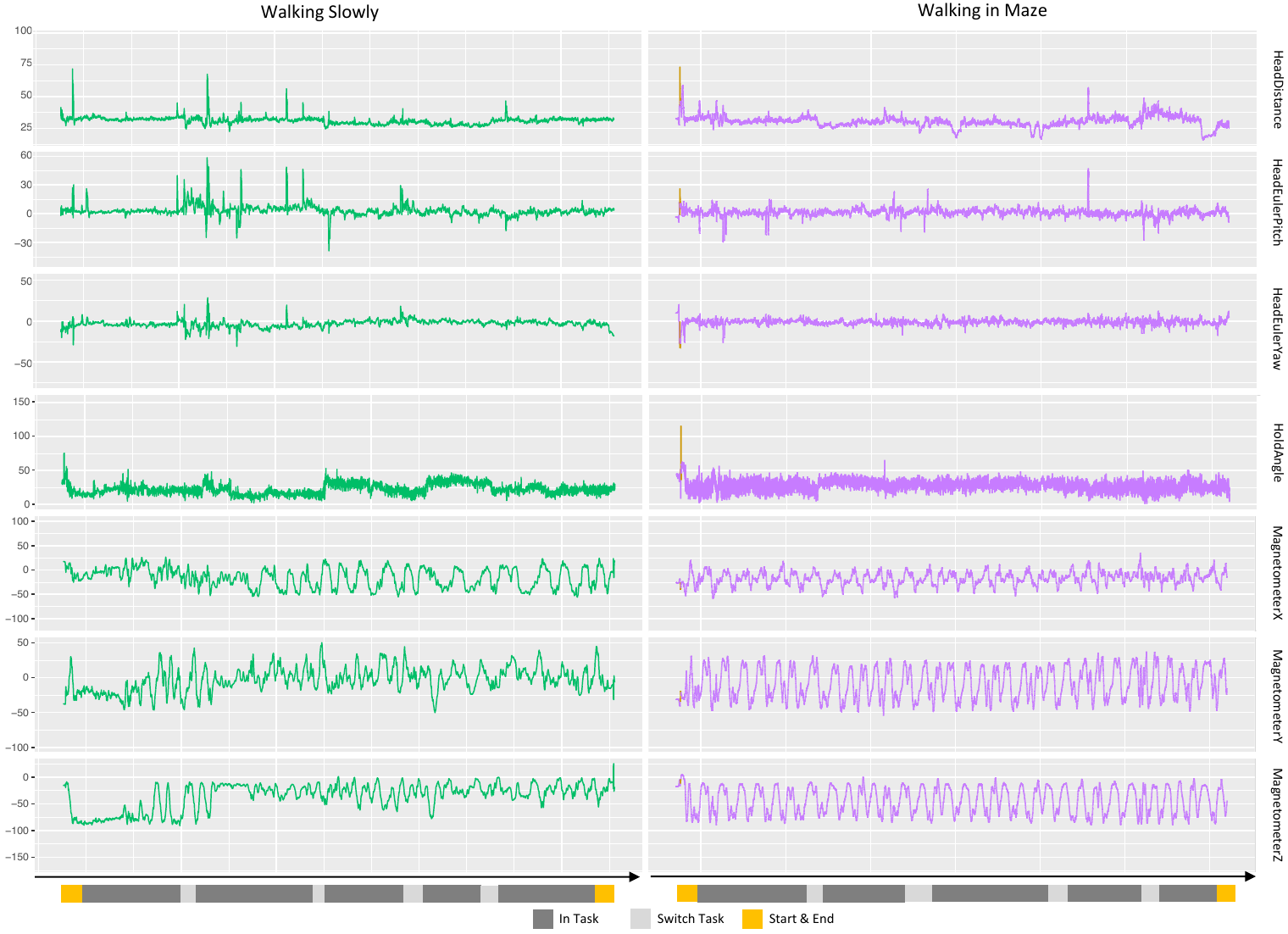}
    \caption{GMM clustering mapped to time-series sensor data on Participant 1 under dynamic conditions}
    \label{fig:exp1gmmtimep1dynamic}
\end{figure}

Figures~\ref{fig:exp1gmmtimep1static} and~\ref{fig:exp1gmmtimep1dynamic} illustrate this pattern for Participant~1 by mapping GMM clusters back onto the time series. In static conditions, clusters align more visibly with task periods and cue-driven transitions. In dynamic conditions, locomotion dominates the signal and makes task-transition boundaries less separable. This motivates a direct comparison between stable in-task periods and task-switch periods.

\noindent\textbf{In vs. Switch Task} We therefore separate each trial into \emph{in-task} periods, where participants were actively engaged in a prompted task, and \emph{switch-task} periods, where they moved between task prompts. This distinction is important because a gaze calibrator may be challenged not only by broad motion condition, but also by brief reorientation moments when the user changes posture, gaze target, or device position.

\begin{table}[!htbp]
\centering
\begin{tabular}{l|cccccccc}
\hline
\multirow{3}{*}{\diagbox{Attribute}{Setting}} & \multicolumn{4}{c}{Lying} & \multicolumn{4}{c}{Sitting} \\
 & \multicolumn{2}{c}{In Task} & \multicolumn{2}{c}{Switch Task} & \multicolumn{2}{c}{In Task} & \multicolumn{2}{c}{Switch Task} \\
 \cline{2-9}
 & Mean & Std & Mean & Std & Mean & Std & Mean & Std \\
  \hline
headDistance & 23.65 & {\cellcolor[rgb]{0.608,0.808,0.494}}2.23 & 31.54 & {\cellcolor[rgb]{0.973,0.412,0.42}}10.10 & 41.25 & {\cellcolor[rgb]{0.388,0.745,0.482}}0.48 & 34.60 & {\cellcolor[rgb]{0.984,0.596,0.455}}8.40 \\
headEulerAnglePitch & 3.96 & {\cellcolor[rgb]{0.388,0.745,0.482}}1.55 & 1.22 & {\cellcolor[rgb]{0.98,0.537,0.447}}9.07 & 4.57 & {\cellcolor[rgb]{0.533,0.784,0.49}}2.31 & 2.19 & {\cellcolor[rgb]{0.988,0.667,0.471}}7.62 \\
headEulerAngleYaw & 0.60 & {\cellcolor[rgb]{0.435,0.757,0.482}}1.80 & -1.35 & {\cellcolor[rgb]{0.973,0.412,0.42}}10.48 & -3.42 & {\cellcolor[rgb]{0.576,0.796,0.49}}2.53 & -0.79 & {\cellcolor[rgb]{0.992,0.729,0.482}}6.90 \\
useraccelerometerX & 0.00 & {\cellcolor[rgb]{0.463,0.765,0.486}}0.08 & 0.01 & {\cellcolor[rgb]{1,0.906,0.518}}0.49 & 0.00 & {\cellcolor[rgb]{0.686,0.827,0.498}}0.19 & 0.01 & {\cellcolor[rgb]{0.949,0.906,0.514}}0.31 \\
useraccelerometerY & 0.00 & {\cellcolor[rgb]{0.388,0.745,0.482}}0.05 & -0.05 & {\cellcolor[rgb]{1,0.918,0.518}}0.38 & 0.05 & {\cellcolor[rgb]{0.463,0.765,0.486}}0.08 & -0.03 & {\cellcolor[rgb]{0.824,0.871,0.506}}0.26 \\
useraccelerometerZ & -0.01 & {\cellcolor[rgb]{0.525,0.784,0.49}}0.11 & 0.01 & {\cellcolor[rgb]{1,0.902,0.514}}0.56 & -0.05 & {\cellcolor[rgb]{0.792,0.859,0.502}}0.24 & -0.01 & {\cellcolor[rgb]{1,0.918,0.518}}0.40 \\
gyroscopeX & 0.00 & {\cellcolor[rgb]{0.388,0.745,0.482}}0.04 & 0.00 & {\cellcolor[rgb]{0.973,0.412,0.42}}0.25 & 0.10 & {\cellcolor[rgb]{0.718,0.839,0.498}}0.09 & 0.00 & {\cellcolor[rgb]{0.992,0.71,0.478}}0.19 \\
gyroscopeY & 0.00 & {\cellcolor[rgb]{0.388,0.745,0.482}}0.04 & 0.00 & {\cellcolor[rgb]{0.973,0.412,0.42}}0.26 & 0.00 & {\cellcolor[rgb]{1,0.894,0.514}}0.19 & 0.00 & {\cellcolor[rgb]{0.98,0.914,0.514}}0.18 \\
gyroscopeZ & 0.00 & {\cellcolor[rgb]{0.388,0.745,0.482}}0.03 & 0.00 & {\cellcolor[rgb]{0.973,0.412,0.42}}0.22 & -0.10 & {\cellcolor[rgb]{0.773,0.855,0.502}}0.09 & 0.00 & {\cellcolor[rgb]{0.992,0.718,0.478}}0.16 \\
orientationRoll & -0.24 & {\cellcolor[rgb]{0.388,0.745,0.482}}0.11 & -0.15 & {\cellcolor[rgb]{0.973,0.412,0.42}}1.70 & 0.00 & {\cellcolor[rgb]{0.404,0.749,0.482}}0.13 & 0.04 & {\cellcolor[rgb]{0.98,0.525,0.443}}1.51 \\
orientationPitch & 0.29 & {\cellcolor[rgb]{0.388,0.745,0.482}}0.04 & 0.01 & {\cellcolor[rgb]{0.973,0.412,0.42}}1.99 & 1.06 & {\cellcolor[rgb]{1,0.898,0.514}}1.21 & -0.28 & {\cellcolor[rgb]{0.992,0.737,0.482}}1.47 \\
orientationYaw & 0.32 & {\cellcolor[rgb]{0.388,0.745,0.482}}0.04 & 0.75 & {\cellcolor[rgb]{0.976,0.914,0.514}}0.39 & 0.53 & {\cellcolor[rgb]{1,0.918,0.518}}0.41 & 0.74 & {\cellcolor[rgb]{0.973,0.412,0.42}}1.59 \\
HoldAngle & 23.29 & {\cellcolor[rgb]{0.388,0.745,0.482}}4.73 & 84.75 & {\cellcolor[rgb]{0.973,0.412,0.42}}36.85 & 30.89 & {\cellcolor[rgb]{0.471,0.769,0.486}}5.69 & 44.03 & {\cellcolor[rgb]{0.996,0.804,0.498}}17.50 \\
Velocity & 0.03 & {\cellcolor[rgb]{0.388,0.745,0.482}}0.02 & 0.10 & {\cellcolor[rgb]{0.973,0.412,0.42}}0.21 & 0.08 & {\cellcolor[rgb]{1,0.894,0.514}}0.14 & 0.09 & {\cellcolor[rgb]{0.976,0.914,0.514}}0.13 \\
\hline
\end{tabular}
\caption{Comparison of mean and standard deviation of motion data for in-task and switch-task periods under static conditions.}\label{tab:exp1-inswitch-task-static}
\end{table}

% Tables~\ref{tab:exp1-inswitch-task-static} and~\ref{tab:exp1-inswitch-task-dynamic} present a detailed comparison of mean and standard deviation of raw motion data in in- and switch-tasks under static and dynamic conditions. The head distance and orientation sensors such as gyroscope and accelerometer exhibit small variation in the in-task state, suggesting users maintain a relatively stable distance from their device while engaging in a task, regardless of their motion state. 

% Tables~\ref{tab:exp1-inswitch-task-static} and~\ref{tab:exp1-inswitch-task-dynamic} distinguish \emph{absolute head-to-camera distance} from \emph{distance variability}. Absolute viewing distance differs across posture and motion condition, but once a participant is engaged in a task, the \emph{fluctuation} around that condition-specific distance is comparatively small. For example, the standard deviation of head distance during in-task segments ranges from 0.48\,cm to 2.73\,cm across conditions, whereas switch-task segments show substantially larger variability, ranging from 7.42\,cm to 10.10\,cm. This suggests that task transitions are a particularly important source of geometry change for handheld gaze tracking.

Tables~\ref{tab:exp1-inswitch-task-static} and~\ref{tab:exp1-inswitch-task-dynamic} distinguish \emph{absolute head-to-camera distance} from \emph{distance variability}. Absolute head-to-camera distance differs across posture and motion condition, but once a participant is engaged in a task, the \emph{fluctuation} around that condition-specific distance is comparatively small. For example, the standard deviation of head-to-camera distance during in-task segments ranges from 0.48\,cm to 2.73\,cm across conditions, whereas switch-task segments show substantially larger variability, ranging from 7.42\,cm to 10.10\,cm. This suggests that task transitions are a particularly important source of geometry change for handheld gaze tracking.

To examine the in-task vs.\ switch-task contrast given the \(n=10\) cohort, and to avoid treating the 56 (motion condition\,\(\times\)\,sensor) table cells as independent observations, we aggregated the comparison to the participant level. For each participant under each movement condition, we took the median ratio of switch-task to in-task variability and tested these ten ratios against the null value of one with a one-sided Wilcoxon signed-rank test. For head-to-camera distance, switch-task variability was reliably larger than in-task variability (median ratio \(2.85\times\); \(W=54\), one-sided \(p=.002\); log-ratio \(t(9)=3.74\), \(p=.005\), Cohen's \(d_z=1.18\)). The same participant-level test across the full sensor panel gave a median ratio of \(1.96\times\) (\(W=52\), one-sided \(p=.005\)). Because this test uses one aggregated value per participant, it is immune to correlations among sensors and conditions, and confirms that the elevated geometry variability at task transitions is a participant-level effect rather than an artefact of pooling correlated table cells.

% To examine the in-task vs.\ switch-task contrast more quantitatively given the \(n=10\) cohort, we treated each cell of Tables~\ref{tab:exp1-inswitch-task-static}--\ref{tab:exp1-inswitch-task-dynamic} as a paired observation and ran a one-sided Wilcoxon signed-rank test on the 56 (motion condition\,\(\times\)\,sensor variable) pairs of in-task vs.\ switch-task standard deviations. Switch-task variability was significantly larger than in-task variability (median switch/in-task SD ratio \(2.07\times\), IQR \(1.21\)--\(5.27\); \(W=1580\), one-sided \(p<.001\)); the equivalent scale-invariant paired \(t\)-test on the log-ratios gave \(t(55)=8.25\), \(p<.001\), Cohen's \(d_z=1.10\). Because the 56 pairs share the same 10 participants and partially overlapping sensor channels, we report this test as an omnibus summary of the descriptive pattern across the table rather than as 56 independent comparisons. Read in that scoped sense, the contrast is consistent across sensor channels and conditions, in line with the interpretation that task transitions are associated with elevated geometry variation within this dataset.

\begin{table}[!htbp]
\centering
\begin{tabular}{l|cccccccc}
\hline
\multirow{3}{*}{\diagbox{Attribute}{Setting}} & \multicolumn{4}{c}{Walking Slowly} & \multicolumn{4}{c}{Walking in Maze} \\
 & \multicolumn{2}{c}{In Task} & \multicolumn{2}{c}{Switch Task} & \multicolumn{2}{c}{In Task} & \multicolumn{2}{c}{Switch Task} \\
 \cline{2-9}
 & Mean & Std & Mean & Std & Mean & Std & Mean & Std \\
  \hline
headDistance & 27.94 & {\cellcolor[rgb]{0.388,0.745,0.482}}2.58 & 31.67 & {\cellcolor[rgb]{0.976,0.439,0.427}}7.42 & 31.37 & {\cellcolor[rgb]{0.424,0.753,0.482}}2.73 & 33.12 & {\cellcolor[rgb]{0.973,0.412,0.42}}7.54 \\
headEulerAnglePitch & 0.23 & {\cellcolor[rgb]{0.467,0.765,0.486}}2.86 & 1.45 & {\cellcolor[rgb]{0.984,0.588,0.455}}8.28 & 2.24 & {\cellcolor[rgb]{0.725,0.839,0.498}}3.78 & 2.22 & {\cellcolor[rgb]{0.973,0.412,0.42}}10.14 \\
headEulerAngleYaw & -1.84 & {\cellcolor[rgb]{0.455,0.765,0.486}}2.82 & -0.33 & {\cellcolor[rgb]{0.996,0.851,0.506}}5.51 & 0.97 & {\cellcolor[rgb]{0.78,0.855,0.502}}3.96 & 0.30 & {\cellcolor[rgb]{0.992,0.773,0.49}}6.34 \\
useraccelerometerX & 0.02 & {\cellcolor[rgb]{0.388,0.745,0.482}}0.51 & 0.02 & {\cellcolor[rgb]{0.992,0.918,0.514}}0.60 & -0.02 & {\cellcolor[rgb]{1,0.918,0.518}}0.60 & 0.01 & {\cellcolor[rgb]{0.973,0.412,0.42}}0.72 \\
useraccelerometerY & -0.09 & {\cellcolor[rgb]{0.388,0.745,0.482}}0.52 & -0.11 & {\cellcolor[rgb]{0.914,0.894,0.51}}0.55 & 0.01 & {\cellcolor[rgb]{1,0.898,0.514}}0.57 & -0.14 & {\cellcolor[rgb]{0.973,0.412,0.42}}0.67 \\
useraccelerometerZ & 0.03 & {\cellcolor[rgb]{0.647,0.82,0.494}}0.52 & 0.05 & {\cellcolor[rgb]{0.992,0.769,0.49}}0.62 & -0.04 & {\cellcolor[rgb]{0.388,0.745,0.482}}0.48 & 0.08 & {\cellcolor[rgb]{0.973,0.412,0.42}}0.74 \\
gyroscopeX & 0.00 & {\cellcolor[rgb]{0.388,0.745,0.482}}0.20 & 0.00 & {\cellcolor[rgb]{0.922,0.898,0.51}}0.25 & 0.01 & {\cellcolor[rgb]{0.996,0.8,0.494}}0.26 & 0.00 & {\cellcolor[rgb]{0.973,0.412,0.42}}0.28 \\
gyroscopeY & -0.03 & {\cellcolor[rgb]{0.588,0.8,0.49}}0.34 & -0.02 & {\cellcolor[rgb]{0.388,0.745,0.482}}0.33 & 0.01 & {\cellcolor[rgb]{0.984,0.62,0.463}}0.38 & 0.01 & {\cellcolor[rgb]{0.973,0.412,0.42}}0.40 \\
gyroscopeZ & -0.04 & {\cellcolor[rgb]{0.776,0.855,0.502}}0.42 & -0.01 & {\cellcolor[rgb]{0.388,0.745,0.482}}0.39 & -0.10 & {\cellcolor[rgb]{0.988,0.659,0.467}}0.45 & 0.02 & {\cellcolor[rgb]{0.973,0.412,0.42}}0.47 \\
orientationRoll & -0.02 & {\cellcolor[rgb]{0.388,0.745,0.482}}0.07 & 0.06 & {\cellcolor[rgb]{0.996,0.835,0.502}}0.30 & -0.02 & {\cellcolor[rgb]{0.878,0.886,0.51}}0.22 & 0.03 & {\cellcolor[rgb]{0.973,0.412,0.42}}0.48 \\
orientationPitch & -0.36 & {\cellcolor[rgb]{0.388,0.745,0.482}}1.79 & -0.27 & {\cellcolor[rgb]{0.514,0.78,0.486}}1.80 & -0.70 & {\cellcolor[rgb]{1,0.91,0.518}}1.91 & 0.03 & {\cellcolor[rgb]{0.973,0.412,0.42}}3.88 \\
orientationYaw & 0.46 & {\cellcolor[rgb]{0.388,0.745,0.482}}0.18 & 0.61 & {\cellcolor[rgb]{0.737,0.843,0.502}}0.31 & 0.68 & {\cellcolor[rgb]{0.996,0.835,0.502}}0.51 & 0.60 & {\cellcolor[rgb]{0.973,0.412,0.42}}0.99 \\
HoldAngle & 26.37 & {\cellcolor[rgb]{0.494,0.773,0.486}}10.34 & 35.22 & {\cellcolor[rgb]{0.973,0.412,0.42}}18.43 & 39.04 & {\cellcolor[rgb]{0.388,0.745,0.482}}9.62 & 34.62 & {\cellcolor[rgb]{0.984,0.561,0.451}}17.07 \\
Velocity & 0.22 & {\cellcolor[rgb]{0.388,0.745,0.482}}0.13 & 0.23 & {\cellcolor[rgb]{0.984,0.576,0.455}}0.17 & 0.08 & {\cellcolor[rgb]{0.788,0.859,0.502}}0.15 & 0.29 & {\cellcolor[rgb]{0.973,0.412,0.42}}0.18 \\
\hline
\end{tabular}
\caption{Comparison of mean and standard deviation of motion data for in-task and switch-task periods under dynamic conditions.}\label{tab:exp1-inswitch-task-dynamic}
\end{table}

The head-distance rows make this contrast concrete. Switch-task periods showed larger head-to-camera distance variability than in-task periods in all four motion conditions: lying (\(31.54 \pm 10.10\)cm), sitting (\(34.60 \pm 8.40\)cm), walking slowly (\(31.67 \pm 7.42\)cm), and walking in maze (\(33.12 \pm 7.54\)cm). These values indicate that task transitions introduce substantial geometry change even when the participant is not walking. This locates a concrete moment at which calibration-based handheld gaze estimation is likely to be challenged: not only during locomotion, but also during the brief reorientation periods between tasks, when users change posture, glance away, or reposition the device.

Taken together, User Study~1 shows that handheld geometry is condition-sensitive and task-phase-sensitive. Sustained task engagement often leads to relatively stable local geometry, whereas transitions between tasks produce larger shifts in head-to-camera distance, head pose, and device orientation. User Study~2 uses this behavioural backdrop to examine how such geometry changes affect calibration-based 2D gaze estimation.

\subsection{User Study 2 -- Impact and Factors}
User Study~2 links the behavioural patterns observed above to gaze-estimation performance. As described in Section~\ref{subsec:impact_factor}, each participant completed five calibration-test pairs per motion condition.
% producing synchronised ground-truth targets, facial images, gaze predictions, and motion signals. 
The analyses below investigate how calibration regime, motion condition, and calibration-test mismatch affect RMSE, and which motion-related variables are most closely associated with error.
We organise the analysis into three steps: (1) \textit{motion data statistics} to examine whether Study~2 elicited comparable motion signatures and whether calibration and test phases differ; (2) \textit{performance degradation} to compare base-model, one-off, task-specific, and motion-specific calibration regimes; and (3) \textit{impact factor identification} to use regression-based analysis to identify which motion-feature groups are most associated with gaze-estimation error.

\subsubsection{Motion Data Visualisation}

\begin{figure}[!htbp]
    \centering
    \includegraphics[width=0.80\textwidth]{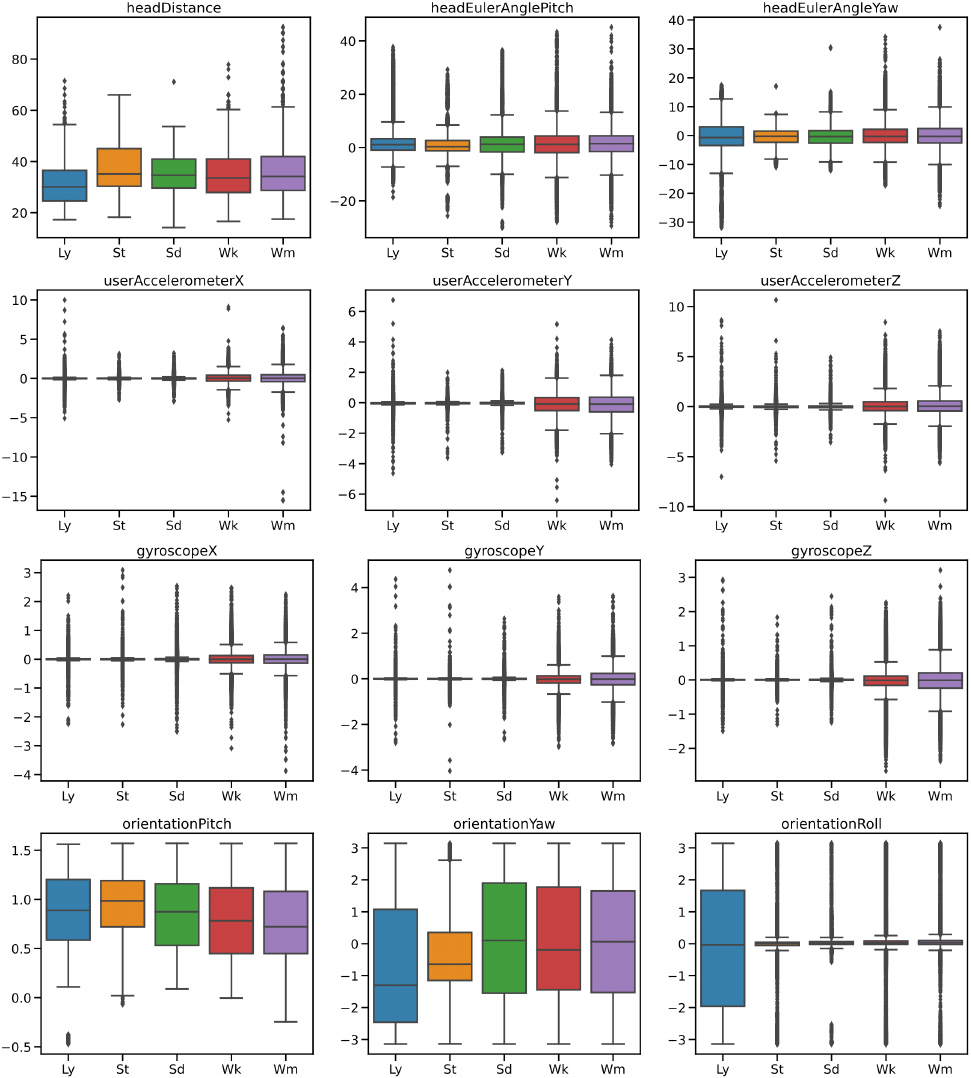}
    \caption{Box plots of motion data under each motion condition: \textit{Ly} for lying, \textit{St}  for sitting, \textit{Sd}  for standing, \textit{Wk} for walking and \textit{Wm} for walking in maze. }
    \label{fig:overall-US3-sensor-data}
\end{figure}

To connect User Study~2 with the behavioural patterns from User Study~1, Figure~\ref{fig:overall-US3-sensor-data} summarises the raw motion data across the five evaluation conditions. We then use Figure~\ref{fig:overall-US3-train-test-motion} to compare calibration and test feature distributions, because calibration-test mismatch is central to the later performance analysis.

\noindent\textbf{Head distance:} The head-to-camera distance pattern broadly matches User Study~1: lying produced the shortest average distance and sitting the longest. In User Study~2, the two dynamic conditions also showed substantial variation, with walking ($M=34.51$\,cm, $SD=8.77$\,cm) and walking in maze ($M=35.52$\,cm, $SD=9.72$\,cm) approaching the variability observed in sitting ($M=37.62$\,cm, $SD=9.47$\,cm). This indicates that the calibration-and-test protocol constrained behaviour but did not eliminate condition-level differences in viewing geometry.

\noindent\textbf{Head Euler Angles (Pitch, Yaw):}  Head-pose means were also consistent with User Study~1: pitch remained positive across conditions and yaw stayed close to 0: lying ($M_{pitch} = 1.21$, $SD_{pitch} = 4.75$, $M_{yaw} = -0.86$, $SD_{yaw} = 6.84$), sitting ($M_{pitch} = 0.95$, $SD_{pitch} = 4.16$, $M_{yaw} = -0.44$, $SD_{yaw} = 2.89$), standing ($M_{pitch} = 1.24$, $SD_{pitch} = 5.34$, $M_{yaw} = -0.38$, $SD_{yaw} = 3.00$), walking ($M_{pitch} = 1.11$, $SD_{pitch} = 5.67$, $M_{yaw} = -0.01$, $SD_{yaw} = 3.93$), and walking in maze ($M_{pitch} = 1.25$, $SD_{pitch} = 0.01$, $M_{yaw} = 0.35$, $SD_{yaw} = 4.69$). However, the variation in pitch and yaw was smaller in User Study~2, e.g., the range of $SD_{pitch}$ is between 4.16 and 5.78, while that in user study 1 is between 7.68 and 10.72. A plausible explanation is that the explicit calibration and 9-point test encouraged participants to stabilise their head and device posture, while the direct use of the experimental phone avoided the extra weight introduced by attaching two devices and longer experiment duration in User Study~1.

\noindent\textbf{User accelerometer and gyroscope:} Mean accelerometer and gyroscope readings remained close to 0, again indicating that the main behavioural differences appear in variability rather than shifted central tendency. Across the two studies, standard-deviation differences were modest for user acceleration and gyroscope streams, less than 0.45 and 0.34, suggesting that the studies elicited broadly comparable low-intensity handheld movement.

\noindent\textbf{Orientation (Pitch, Yaw, Roll):} Device orientation differed more visibly between the two studies. In User Study~2, participants often adjusted their phone orientation to complete calibration and target fixation, especially in lying and standing conditions. That is, lying ($M_{Pitch} = 0.87$, $SD_{Pitch} = 0.40$, $M_{Yaw} = -0.72$, $SD_{Yaw} = 1.93$, $M_{Roll} = -0.16$, $SD_{Roll} = 2.04$),  sitting ($M_{Pitch} = 0.94$, $SD_{Pitch} = 0.33$, $M_{Yaw} = -0.39$, $SD_{Yaw} = 1.18$), $M_{Roll} = -0.18$, $SD_{Roll} = 0.87$), standing ($M_{Pitch} = 0.85$, $SD_{Pitch} = 0.37$, $M_{Yaw} = 0.13$, $SD_{Yaw} = 2.01$), $M_{Roll} = -0.01$, $SD_{Roll} = 0.45$), walking ($M_{Pitch} = 0.80$, $SD_{Pitch} = 0.39$, $M_{Yaw} = -0.05$, $SD_{Yaw} = 1.78$), $M_{Roll} = 0.09$, $SD_{Roll} = 0.78$), and walking in maze ($M_{Pitch} = 0.77$, $SD_{Pitch} = 0.08$, $M_{Yaw} = 0.03$, $SD_{Yaw} = 1.78$), $M_{Roll} = 0.08$, $SD_{Roll} = 0.66$). This reinforces the need to analyse calibration and test phases directly rather than assuming that a motion condition has a single stable geometry. 

\begin{figure}[!htbp]
    \begin{subfigure}{0.45\textwidth}
        \centering
    \includegraphics[width=\textwidth, height = 140pt]{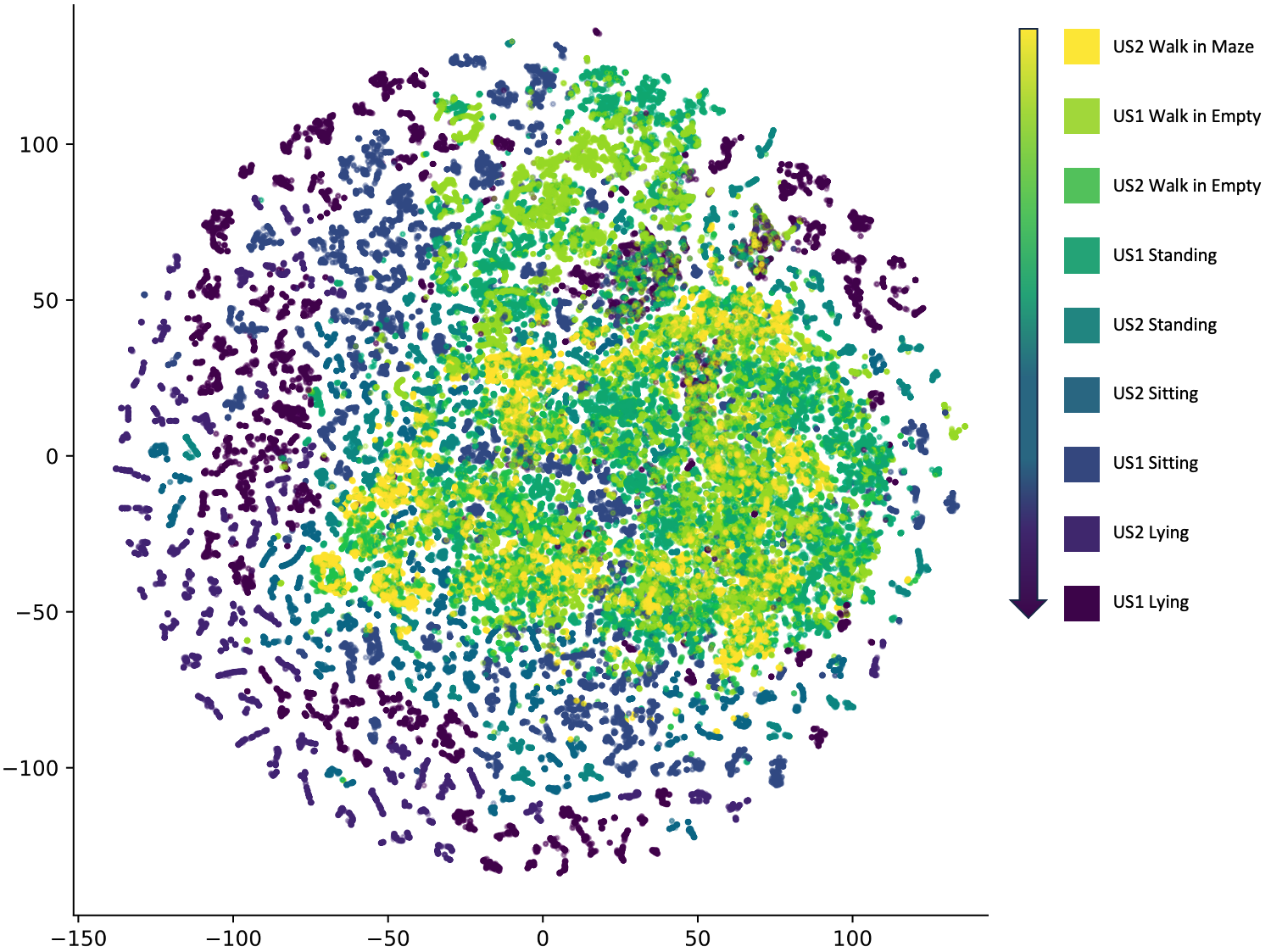}
    \caption{Overall comparison}
    \label{fig:exp13-tsne-all}
    \end{subfigure}
    \begin{subfigure}{0.52\textwidth}
        \centering
         \includegraphics[width=\textwidth, height = 140pt]{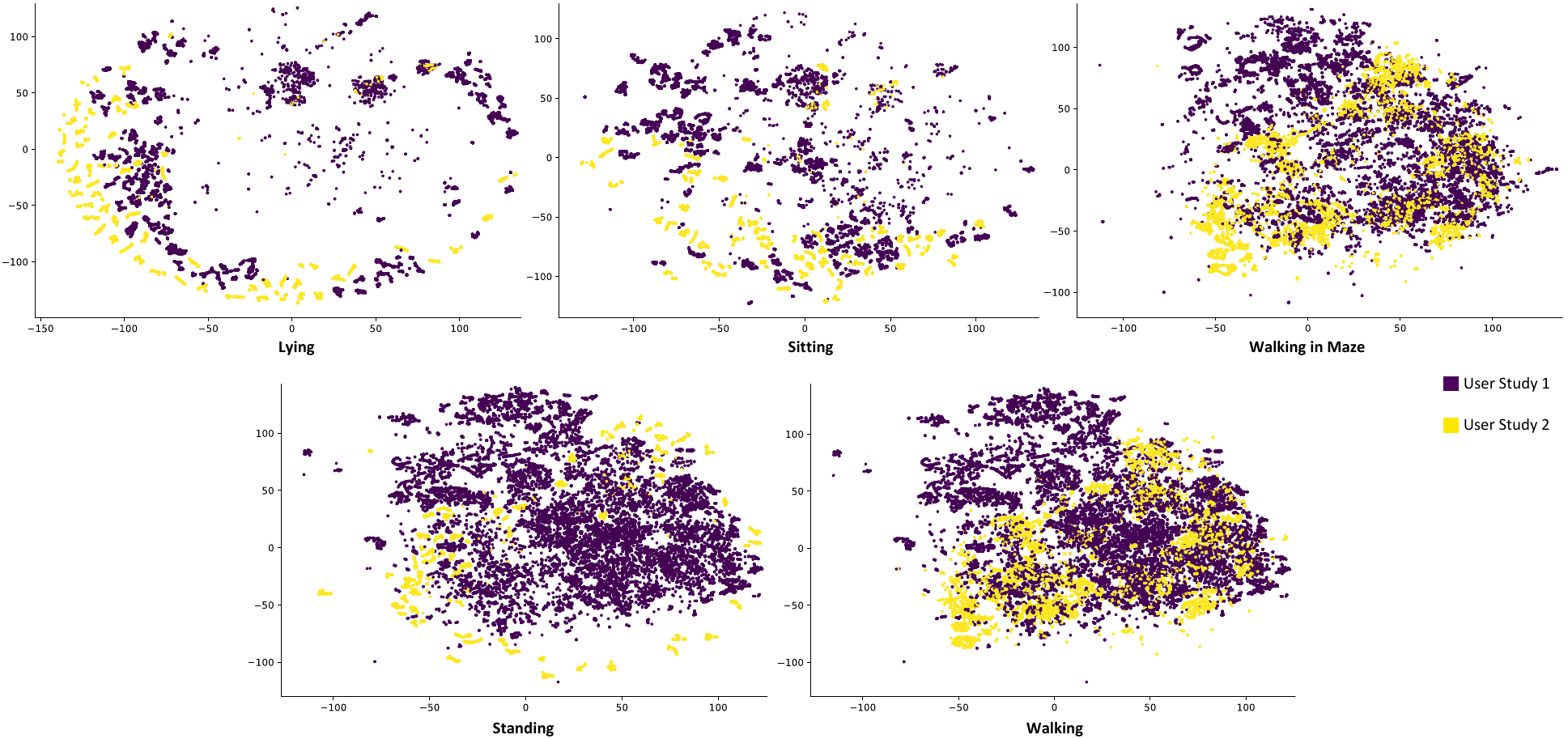}
    \caption{Comparison on each of motion conditions}
    \label{fig:exp13-tsne-posture}
    \end{subfigure}
    \caption{Cross-comparison t-SNE plots for User Study 1 and 2}
    \label{fig:user-study-1-2-comparison}
\end{figure}

Figure~\ref{fig:user-study-1-2-comparison} compares the feature distributions across the two studies. The left panel gives an aggregate overview, while the right panel compares conditions, with the walking conditions combined for this cross-study view. The two studies show qualitative overlap in feature distributions, indicating broadly comparable geometry patterns at an aggregate level. We treat this as a qualitative similarity check rather than as evidence of behavioural equivalence, since Study~2 also included calibration and 9-point testing.

\subsubsection{Impact of Mobility on 2D Gaze Estimation}
\begin{figure}[!htbp]
    \centering
    \includegraphics[width=0.95\textwidth]{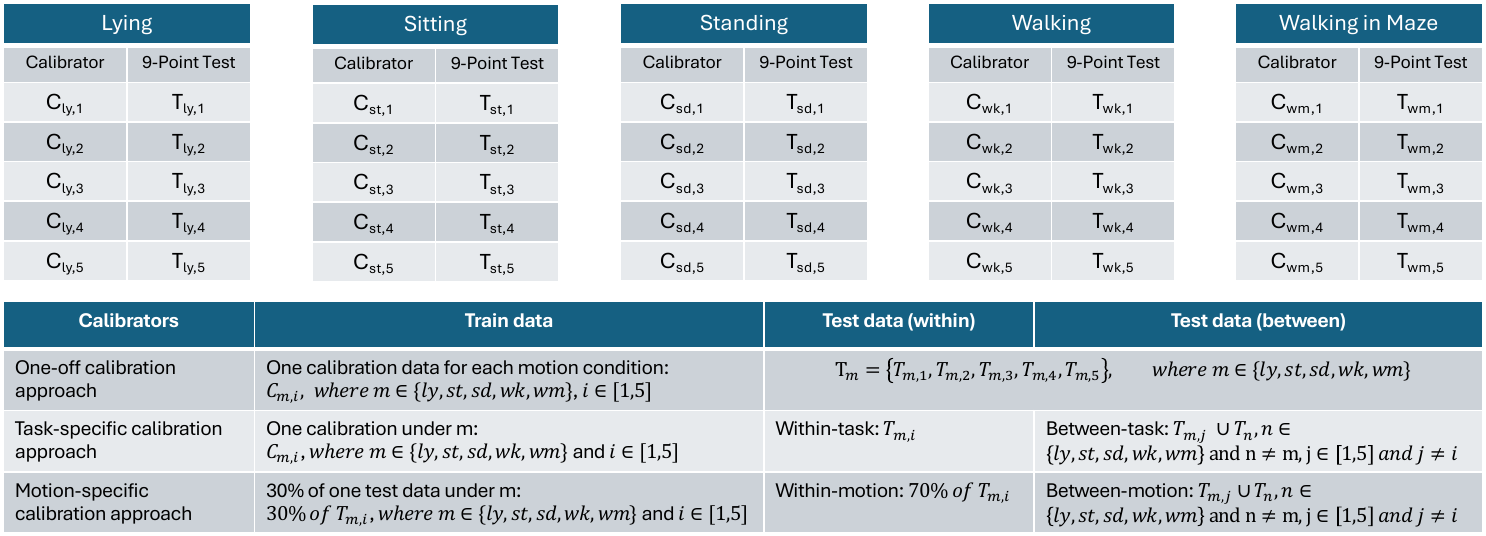}
    % \caption{Person-specific training and test regime of calibrations in different granularities for each motion condition}
    \caption{Overview of participant-specific calibration and test regimes at different behavioural granularities.}
    \label{fig:train-test-calibrators}
\end{figure}

We next evaluate how the 2D gaze-estimation pipeline performs under different calibration regimes. Figure~\ref{fig:train-test-calibrators} summarises the person-specific training and testing procedures, moving from an uncalibrated base model to increasingly behaviour-matched calibration settings.

We first compare two baseline settings: (1) \textit{base model} uses the iTracker-style model described in Section~\ref{subsec:eye_tracking_module} without participant-specific calibration; (2) \textit{one-off calibration} setting trains a single participant-specific calibrator using one calibration sequence and then applies it across the corresponding test data, approximating the common practice of calibrating once before use~\cite{Tobii2023eyetrackerusage, Gazepoint2022tutorial}.

% First we conduct an in-depth analysis by comparing the gaze estimation performance on two settings: (1) \textit{base model performance}; that is, directly inputting the facial landmarks and facial images to the iTracker model in Section~\ref{subsec:eye_tracking_module} and obtaining the gaze points; (2) \textit{one-off calibration approach}; that is, training one calibrator using any of the calibration data following the procedure in Section~\ref{subsec:calibration_module} and then using this calibrator for gaze point prediction. This references the default procedure of commercial eye-trackers~\cite{Tobii2023eyetrackerusage, Gazepoint2022tutorial}, where a calibration is performed once before using the eye-tracker, and then the same calibrator is applied until the entire experiment is completed.

The 2D gaze estimation performance measured by calculating the Euclidean distance error in centimetres (cm), i.e. Root Mean Squared Error (RMSE), to quantify the discrepancy between the predicted gaze points and the ground truth: $\text{RMSE} = \sqrt{\frac{1}{n} \sum_{i=1}^n (gt_i - \hat{p}_i)^2}$, where $gt_i$ and $\hat{p}_i$ are the $i$th ground truth and estimated points and $n$ is the total number of samples. Table~\ref{tab:exp3overallsystem} reports participant-averaged RMSE for the base model and one-off calibration. For the one-off setting, each of the 5 calibration sequences was used once to train a calibrator, tested on the 5 test sequences, and then averaged.

% We first compare two baseline settings. The \textit{base model} uses the iTracker-style model described in Section~\ref{subsec:eye_tracking_module} without participant-specific calibration. The \textit{one-off calibration} setting trains a single participant-specific calibrator using one calibration sequence and then applies it across the corresponding test data, approximating the common practice of calibrating once before use~\cite{Tobii2023eyetrackerusage, Gazepoint2022tutorial}.

Within the present sample (n=10), we conducted a repeated-measures ANOVA to examine the effect of motion condition (\textit{lying, sitting, standing, walking, walking in maze}) on the per-participant mean RMSE of the one-off calibrator. Mauchly's test indicated that the sphericity assumption was met ($\chi^2(9) = 13.76$, $p = .140$). The ANOVA showed that the main effect of motion condition on overall RMSE \emph{did not reach statistical significance} ($F(4,36) = 0.85$, $p = .506$, $\eta_p^2 = .018$). Holm--Bonferroni-corrected pairwise comparisons likewise yielded no significant pairwise differences (all corrected $p \geq .38$); the largest descriptive contrast was between Walking and Walking-in-maze ($t(9) = -2.43$, $p_\text{unc} = .038$, Cohen's $d = -0.22$, not surviving correction). We report these effect sizes as \emph{exploratory within-sample estimates}: with the present participant pool the data do not support a strong claim of a global main effect of motion on aggregate RMSE under the one-off regime. As we show below, motion-related effects emerge more clearly when the analysis is conditioned on the calibration regime and on the calibration--test behavioural mismatch.

\begin{table}[!htbp]
\centering
\resizebox{\textwidth}{!}{%
\begin{tabular}{l|cccccc} 
\hline
\diagbox{System}{RMSE (cm)}{Scenario} & Average & Lying & Sitting & Standing & Walking & Walking in Maze  \\ 
\hline
Base Model & 4.49 (2.76) & 4.13 (2.45) & 4.41 (2.93) & 4.48 (2.77) & 4.54 (2.64) & 4.88 (2.88)\\
One-off Calibration & 3.44 (2.19) & 3.53 (2.18) & 3.07 (1.97) & 3.24 (2.05) & 3.56 (2.29) & 3.79 (2.44)\\
\hline
\end{tabular}
}
% \caption{Comparison of RMSE (mean and std in brackets) for base model and one-off calibration} \label{tab:exp3overallsystem}
\caption{Comparison of RMSE for the base model and one-off calibration. Values are mean (standard deviation) in cm.} \label{tab:exp3overallsystem}
\end{table}

\noindent\textbf{Base model performance}. The base model provides a benchmark for uncalibrated use. It was pre-trained on third-party mobile gaze data and fine-tuned on device-specific RGBDGaze data~\cite{arakawa2022rgbdgaze}, but it was not adapted to the current participants. As shown in Table~\ref{tab:exp3overallsystem}, average RMSE was 4.49\,cm, with the largest descriptive error in walking in maze (4.88\,cm).
% Therefore, it indicates that the base model, without additional calibration, is impractical for an eye-tracking system, particularly under dynamic conditions.
% Therefore, in the present setup, the base model without user-specific calibration appears insufficient for accurate point-level handheld gaze interaction, especially under more dynamic conditions. The one-off calibration approach improves performance, but the remaining error suggests that a single short calibration does not fully cover the geometry variation encountered even within this semi-controlled study.

Therefore, in the present setup, the base model without user-specific calibration appears insufficient for accurate point-level handheld gaze interaction. The one-off calibration approach improves performance, but the remaining error suggests that a single short calibration does not fully cover the geometry variation encountered even within this semi-controlled study.

\noindent\textbf{One-off calibration approach performance}. 
 
One-off calibration reduced average RMSE from 4.49\,cm to 3.44\,cm, showing the value of participant-specific adjustment. Descriptively, sitting had the lowest error (3.07\,cm) and walking in maze the highest error (3.79\,cm). However, consistent with the ANOVA above, these condition differences should be read as descriptive rather than as statistically reliable global differences under the one-off regime. The main result is that one-off calibration improves the base model but leaves several centimetres of residual error.

\begin{figure}[!htbp]
    \centering
    \includegraphics[width=0.7\textwidth]{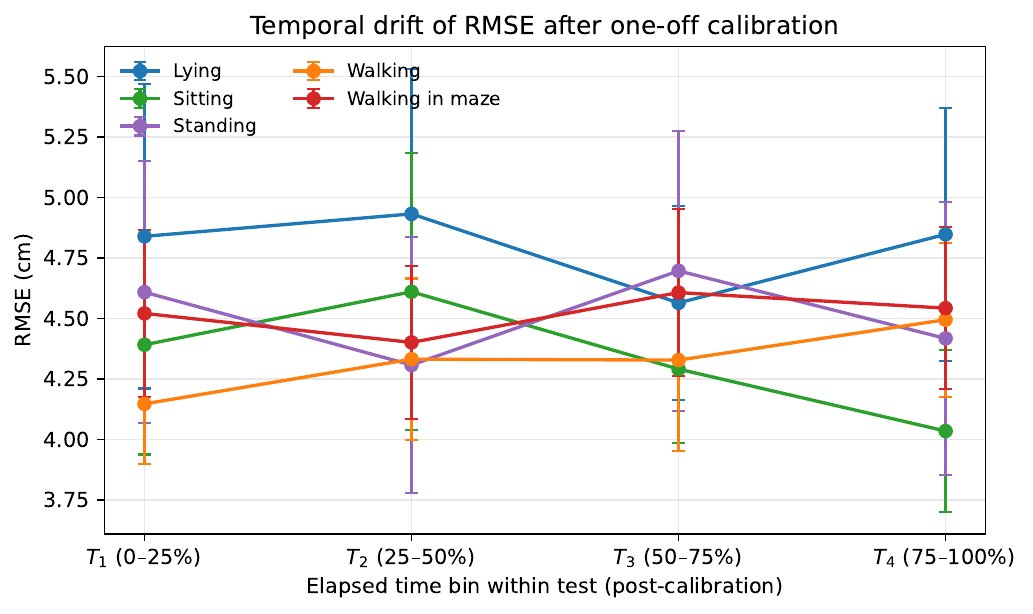}
    \caption{Held-out RMSE as a function of elapsed time within the test sequence (four equal temporal bins after one-off calibration), per motion condition. Error bars indicate $\pm 1$ standard error across participants. Within the present dataset there is no monotonic post-calibration drift across the four bins.}
    \label{fig:temporal-drift}
\end{figure}

\noindent\textbf{Temporal drift after one-off calibration.}
To examine whether one-off calibration degrades over time, we additionally analysed the held-out RMSE as a function of elapsed time since calibration. For each calibration--test pair (with calibrator trained on a different test sequence than the one used for evaluation), we partitioned the post-calibration test samples into four equal temporal bins: $T_1$ (0--25\%), $T_2$ (25--50\%), $T_3$ (50--75\%), and $T_4$ (75--100\%) of elapsed time within the test sequence. The empirical pattern is shown in Figure~\ref{fig:temporal-drift}. Across all motion conditions, RMSE remained largely flat across bins, moving from approximately 4.50\,cm in $T_1$ to 4.47\,cm in $T_4$, with no monotonic increase. A two-way repeated-measures ANOVA on temporal bin and motion condition showed neither a significant main effect of elapsed time ($F(3,27)=0.05$, $p_\text{GG}=.988$, $\eta_p^2 < .001$) nor a significant interaction with motion condition ($F(12,108)=1.03$, $p_\text{GG}=.509$, $\eta_p^2=.011$). \emph{Within this dataset and the present time scale (a single 9-point test sequence), we therefore did not find evidence of systematic post-calibration drift}; the dominant source of variability appears instead to be condition-level mismatch between calibration and test, not within-sequence temporal accumulation.

\noindent\textbf{Task-specific calibration approach performance}. 
We then evaluated whether calibration improves when the calibration data better match the task context at test time. In the task-specific setting, each calibration sequence trains a calibrator that is tested both on its corresponding task test sequence (\textit{within-task}) and on the other task test sequences from the same motion condition (\textit{between-task}).

\begin{figure}[!htbp]
    \centering
    \includegraphics[width=0.65\textwidth]{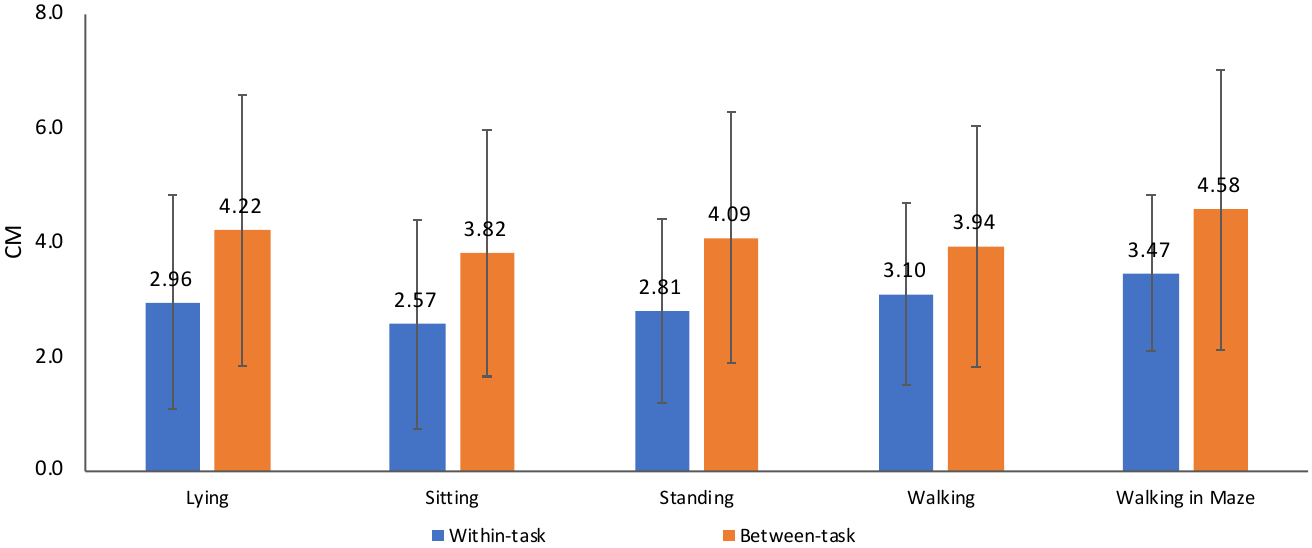}
    % \caption{Comparison of RMSE for within-task and between-task scenarios}
    \caption{Within-task and between-task RMSE under task-specific calibration.}
    \label{fig:within-between}
\end{figure}

For each motion condition, we iterated this procedure across the five calibration sequences and averaged the resulting RMSE values. Figure~\ref{fig:within-between} reports the within-task and between-task results.

% On average, the improvement is 0.45 cm on all the conditions; i.e., 13.41\% from the one-ff calibration approach performance. In the sitting condition, the error is close to the test error on GazeCapture dataset; i.e., 2.05 cm. However, most of the conditions have not yet achieved this accuracy, due to the natural difference in postures during calibration and testing. Figure~\ref{fig:overall-US3-train-test-motion} presents t-SNE plots of calibration and test data on each motion condition aggregated on all the participants. There is distribution shift in the motion data between calibration and testing stage indicate head and body behaviour differences between the two phases.
Compared with one-off calibration, the task-specific calibration regime reduced error across all motion conditions, with an average improvement of approximately 0.45\,cm (13.41\%) in the present dataset. The largest descriptive gain was observed in \emph{sitting} ($-$0.50\,cm), while the smallest was observed in \emph{walking in maze} ($-$0.32\,cm). This pattern suggests that calibrators benefit when the behavioural regime at test more closely matches the regime sampled during calibration.

\begin{figure}[!htbp]
    \centering
    \includegraphics[width=0.85\textwidth]{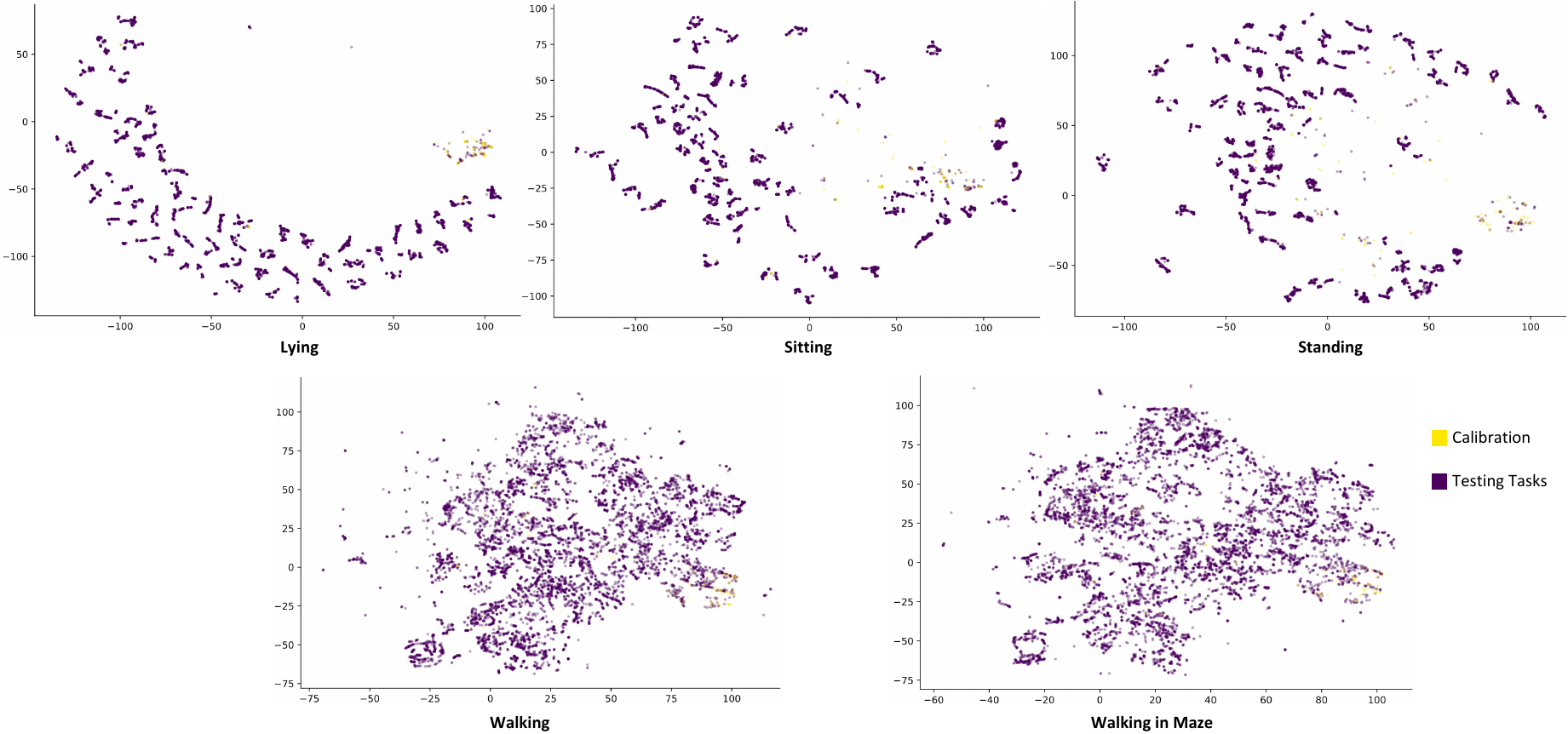}
    % \caption{t-SNE plots of motion data of calibration and test dataset under each motion condition}
    \caption{t-SNE visualisation of calibration and test motion-feature distributions under each motion condition.}
    \label{fig:overall-US3-train-test-motion}
\end{figure}

Figure~\ref{fig:overall-US3-train-test-motion} further supports this interpretation. The calibration and test feature distributions are not fully overlapping, indicating a calibration-test distribution shift in head and body behaviour. This mismatch provides a plausible explanation for why even a short within-condition calibration does not fully eliminate error.

\noindent\textbf{Motion-specific calibration approach performance.} 
Finally, we evaluated a motion-specific calibration setting, the most behaviour-matched regime available in the present dataset. For each motion condition, 30\% of the 9-point data from that condition was used to train a calibrator and the remaining 70\% was used for \textit{within-motion} testing. The same calibrator was also evaluated on data from the other motion conditions to estimate \textit{between-motion} transfer. This process was repeated 5 times per motion condition and averaged.

% The motion-specific calibration approach achieved high accuracy, with errors of less than 2 cm in stable conditions, but under dynamic conditions the error increases to 2.73 cm, dropping by nearly 48.91\%. This represents a 36.15\% reduction in error compared to the one-off calibration approach. These results suggest an upper limit of performance in our study and they indicate that the model can greatly improve its performance by appreciating more body and head movements.

The motion-specific calibration approach achieved the best performance among the calibration regimes considered here. In static conditions, within-motion error was below 2\,cm for \emph{lying, sitting, and standing} (1.78--1.88\,cm), whereas dynamic conditions remained more difficult, reaching approximately 2.54\,cm in walking and 2.91\,cm in walking in maze. Relative to the one-off calibration regime, the motion-specific regime reduced error by approximately 36.15\% on average across conditions. However, the remaining gap between static and dynamic conditions indicates that even richer within-condition calibration data does not fully remove the effect of motion-related geometry change.

These numbers should be read as an upper bound for the present semi-controlled evaluation rather than as performance in unconstrained mobile use; we return to this scope in Section~\ref{sec:discussion}.

\begin{figure}[!htbp]
\centering
\includegraphics[width=0.75\textwidth]{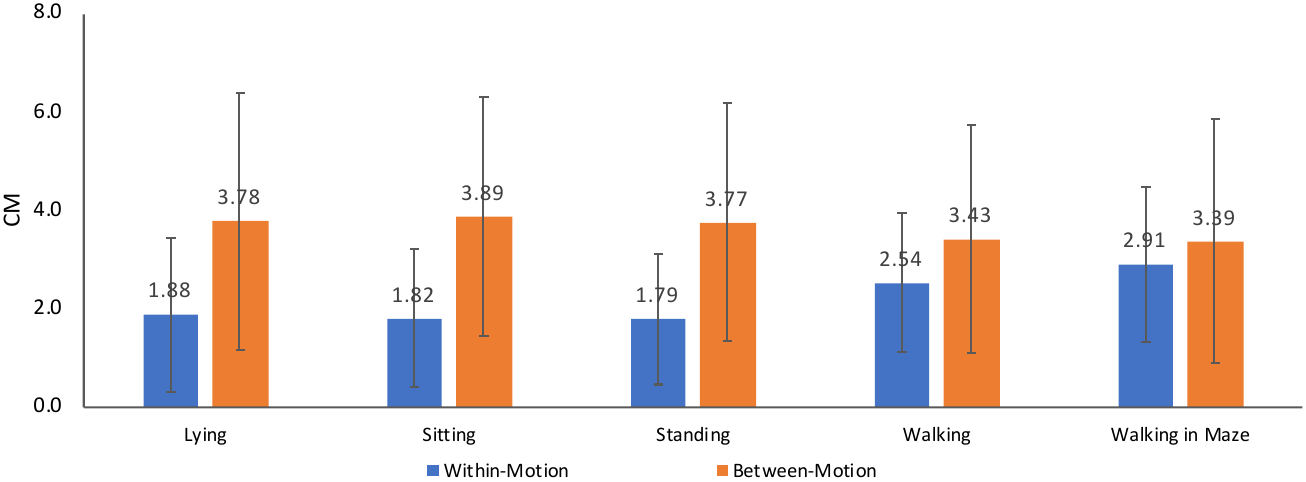}
% \caption{Comparison of RMSE for within-motion and between-motion conditions}
\caption{Within-motion and between-motion RMSE under motion-specific calibration.}
\label{fig:exp3textlevelacc}
\end{figure}

Table~\ref{tab:exp3motionleveltraintest} shows the corresponding train-test matrix. The diagonal entries are consistently the lowest within each test condition, indicating that calibrators perform best when training and testing share the same motion condition. Transfer across conditions increases error even between static postures, such as training on sitting and testing on standing (1.82\,cm to 3.36\,cm). The degradation is larger when transferring between static and dynamic conditions, and is especially visible for lying as a test condition. This pattern supports the interpretation that calibration quality depends on the geometry sampled during calibration, not only on participant identity.

\begin{table}[!htbp]
\centering
{\footnotesize
\begin{tabular}{c|ccccc} 
\hline
\multicolumn{1}{l|}{\diagbox{{}Test}{RMSE (cm)}{{}Train}} & Lying & Sitting & Standing & Walking & Walking in Maze \\
\hline
Lying                     & {\cellcolor[rgb]{0.737,0.843,0.502}}1.88 (1.57)~ & {\cellcolor[rgb]{0.988,0.69,0.475}}3.89 (2.27)~  & {\cellcolor[rgb]{0.992,0.71,0.478}}3.75 (2.11)~  & {\cellcolor[rgb]{1,0.886,0.514}}3.50 (2.10)~     & {\cellcolor[rgb]{0.996,0.816,0.498}}3.59 (2.06)~  \\
Sitting                   & {\cellcolor[rgb]{1,0.875,0.51}}3.52 (2.04)~      & {\cellcolor[rgb]{0.388,0.745,0.482}}1.82 (1.39)~ & {\cellcolor[rgb]{0.922,0.898,0.51}}3.36 (1.96)~  & {\cellcolor[rgb]{0.851,0.878,0.506}}3.29 (1.94)~ & {\cellcolor[rgb]{0.847,0.875,0.506}}3.31 (1.97)~  \\
Standing                  & {\cellcolor[rgb]{0.98,0.914,0.514}}3.49 (2.01)~  & {\cellcolor[rgb]{0.882,0.886,0.51}}3.31 (2.06)~  & {\cellcolor[rgb]{0.518,0.78,0.486}}1.78 (1.32)~  & {\cellcolor[rgb]{0.855,0.878,0.506}}3.19 (2.01)~ & {\cellcolor[rgb]{0.812,0.867,0.506}}3.21 (1.95)~  \\
Walking                   & {\cellcolor[rgb]{0.988,0.643,0.467}}3.89 (2.41)~ & {\cellcolor[rgb]{0.984,0.573,0.451}}4.05 (2.39)~ & {\cellcolor[rgb]{0.992,0.706,0.478}}3.86 (2.34)~ & {\cellcolor[rgb]{0.761,0.851,0.502}}2.54 (1.42)~ & {\cellcolor[rgb]{1,0.922,0.518}}3.46 (2.51)~      \\
Walking in maze           & {\cellcolor[rgb]{0.976,0.463,0.431}}4.22 (2.33)~ & {\cellcolor[rgb]{0.973,0.412,0.42}}4.31 (2.44)~  & {\cellcolor[rgb]{0.976,0.486,0.435}}4.12 (2.43)~ & {\cellcolor[rgb]{0.992,0.714,0.478}}3.73 (2.33)~ & {\cellcolor[rgb]{0.937,0.902,0.514}}2.91 (1.58)~ 
 \\\hline
\end{tabular}}
% \caption{Comparison of RMSE (mean and std in brackets) for between-motion-state experiments}\label{tab:exp3motionleveltraintest}
\caption{Train-test RMSE matrix for the motion-specific calibration setting. Values are mean (standard deviation) in cm.}\label{tab:exp3motionleveltraintest}

\end{table}

\subsubsection{Impact factor identification.} 
The calibration analyses above suggest that gaze error depends on how well the calibration data cover the geometry encountered during testing. We therefore examine which motion-related signals are most associated with RMSE. The goal is interpretability rather than causal attribution: the regression analysis ranks sensor-feature groups that covary with error under the present design.
To do so, we construct a dataset in which extracted sensor features serve as predictors and gaze-estimation error serves as the target. Following the feature extraction procedure in Section~\ref{subsec:feature}, calibrators are trained using 30\% of the test data from each motion condition and evaluated on the remaining within-condition data and on data from other motion conditions.
% Given the high dimensionality of our dataset (552 features across various sensors), we employ the Lasso (Least Absolute Shrinkage and Selection Operator) regression model. Lasso is particularly suited for this analysis as it handles multicollinearity and performs automatic feature selection by assigning zero coefficients to less important features. We use LassoCV from sklearn with cross-validation and optimal alpha parameter selection to enhance the robustness of our results.

% Given the high dimensionality of the dataset (552 features across multiple correlated sensor channels), we selected Lasso as our primary interpretability-oriented model. Our goal in this stage was not to maximise predictive accuracy, but to obtain a sparse and comparatively stable ranking of motion-related factors while reducing multicollinearity. Lasso is suitable for this purpose because the \(L_1\) penalty suppresses weak predictors and encourages a more compact set of non-zero coefficients than ordinary least squares or Ridge regression.

Given the high dimensionality of the dataset (552 features across multiple correlated sensor channels), we selected Lasso as our primary interpretability-oriented model. Our goal in this stage was not to maximise predictive accuracy, but to obtain a sparse ranking of motion-related factors under the present design while reducing coefficient instability from multicollinearity. Lasso is suitable for this purpose because the \(L_1\) penalty suppresses weak predictors and encourages a more compact set of non-zero coefficients than ordinary least squares or Ridge regression.

Table~\ref{tab:exp3RegMobility} reports the top-ranked features for the between-motion and within-motion analyses. At the individual-feature level, the highest-ranked predictors repeatedly come from head-to-camera distance, head-pose, device-orientation, and magnetometer-derived feature families. The main value of this feature list is to reveal recurring geometry-related groups, which we summarise at the factor-group level below.

\begin{table}[!htbp]
\centering
\includegraphics[width=\textwidth]{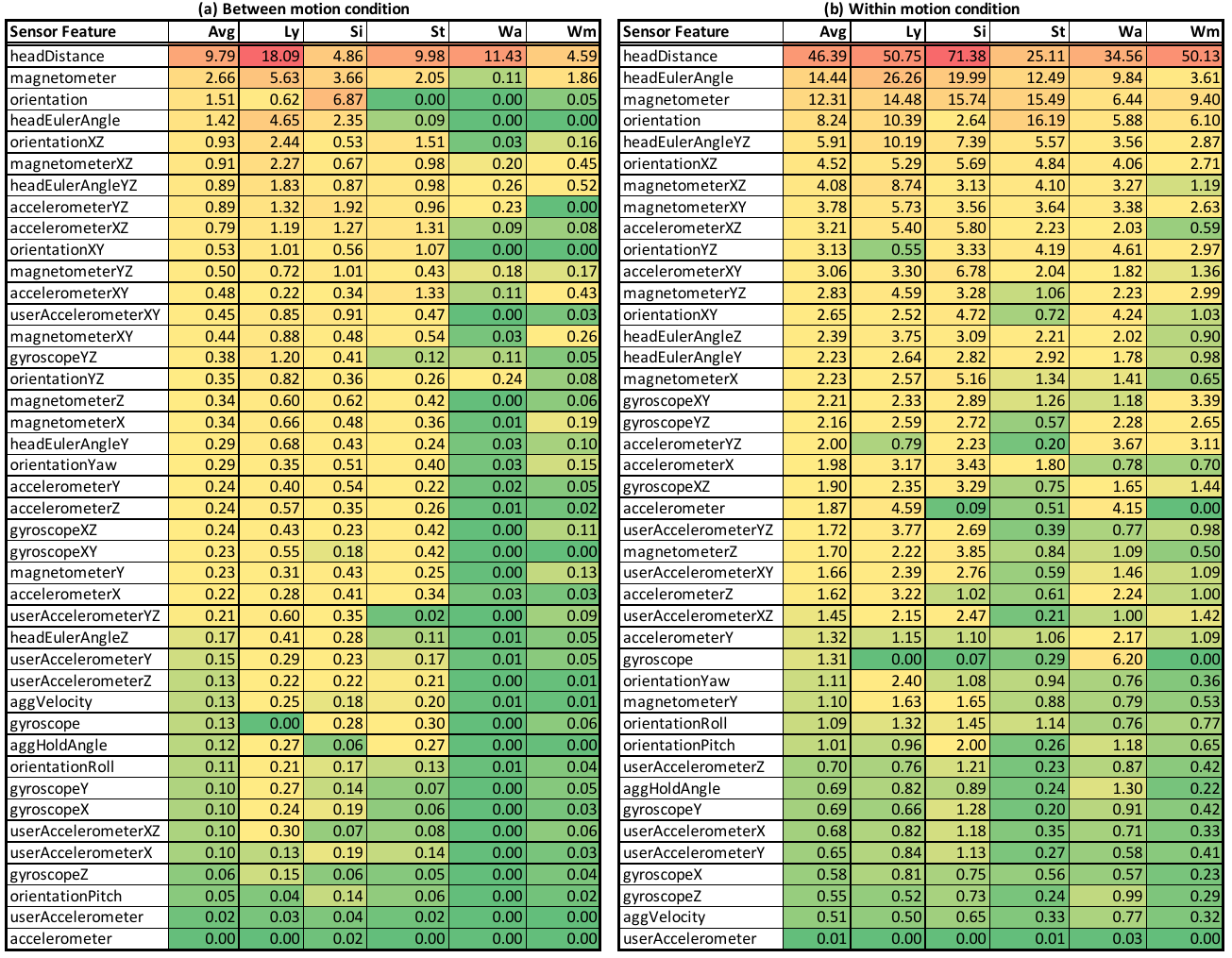}
\caption{Impact factors identified \textit{between} and \textit{within} motion conditions}\label{tab:exp3RegMobility}
\end{table}

\section{Discussion}\label{sec:discussion}
% Our studies offer an empirical analysis of user-device interaction under varying motion states, shedding light on the core limitations of 2D gaze estimation in mobile contexts, particularly regarding head-movement adaptation. By examining both low-level sensor signals (e.g.\ accelerometer, gyroscope) and higher-level behavioural patterns (in-task vs.\ switch-task, within- vs.\ between-motion conditions), we highlight the need for more robust and adaptive eye-tracking systems on handheld devices. Below, we discuss the main takeaways and potential paths forward.

The results refine the general claim that motion matters for handheld 2D gaze estimation. In the representative front-camera RGB 2D pipeline studied here, degradation is structured by \emph{when} geometry changes, \emph{how} calibration samples that geometry, and \emph{which} motion signals dominate in a given condition. Three observations are central to the discussion. First, user-device geometry is comparatively stable during sustained in-task interaction but becomes substantially more variable during task switches. Second, the strongest performance differences appear when calibration and test data are compared by behavioural match, rather than when motion condition is treated only as a global main effect. Third, factor rankings are condition-sensitive: head-to-camera distance, head movement, and device-orientation change recur across analyses, but their relative importance varies by motion condition. These findings position the contribution as a condition-sensitive account of motion-related degradation, rather than a broad claim that all mobility uniformly worsens gaze estimation.

\subsection{Behaviour Patterns in Mobile Interaction}

\subsubsection{Sensor data analysis and insights}
The sensor analysis shows that handheld stability is not determined by locomotion alone. Static postures such as lying and sitting can still produce substantial variation in head-to-camera distance and device orientation because participants have freedom to adjust their arms, hands, and body position. Conversely, walking increases accelerometer and gyroscope variability, but participants appear to constrain their phone posture into fewer recurring patterns while moving. This distinction matters for gaze estimation because the relevant challenge is not simply whether the user is static or mobile, but how stable the face-camera-screen geometry remains during the current interaction segment.

\subsubsection{Clustering distribution and user behaviour}
The t-SNE and GMM analyses support the same interpretation at the feature level. Lying and sitting form more distinguishable regions, while the walking conditions overlap more strongly because locomotion imposes a shared motion signature. The cluster counts further suggest that static conditions allow a wider range of postural configurations, whereas walking leads to fewer recurring handheld patterns. Thus, user behaviour is structured, but the structure differs by posture and movement context.

\subsubsection{In-Task vs. Switch-Task dynamics}
The clearest behavioural contrast is between \emph{in-task} stability and \emph{switch-task} variability. During continuous engagement with a single activity, participants tended to maintain a relatively stable local geometry. During task switches, however, head-to-camera distance, head pose, and device orientation varied much more strongly; head-distance standard deviations reached approximately 7.42--10.10\,cm across conditions. This does not by itself mean that every transition immediately increases gaze error, but it identifies task switching as a concrete moment when a calibration-based system is likely to face geometry outside the recent in-task regime.

% \subsection{Limitations of 2D daze estimation in handling dynamic head poses}
\subsection{Limitations of 2D gaze estimation under dynamic handheld geometry}

% Our study provides empirical evidence that 2D gaze estimation methods, while popular for their simplicity, are particularly vulnerable to changes in head pose. This weakness stems from the fundamental assumption in 2D approaches that the model can implicitly learn to handle head pose variations without explicit normalisation or modelling.

The findings suggest that the representative 2D gaze-estimation pipeline is most challenged when the geometry between face, camera, and screen moves outside the regime sampled during calibration or training. This interpretation is consistent with the design of many 2D appearance-based pipelines, which often absorb geometry variation implicitly through data rather than explicitly normalising it. Importantly, the evidence is not simply a global ``walking is worse'' effect: the repeated-measures ANOVA on one-off calibration did not show a statistically reliable main effect of motion condition, while the calibration-regime and train-test analyses showed clearer degradation when behavioural regimes did not match. Robustness should therefore be judged with respect to calibration-test geometry, not only average benchmark accuracy or a single seated calibration protocol.

% Superseded by the rev2 condition-sensitive itemisation below; previous numbers (70\%, 80\%) reflected an aggregated coefficient sum that was later replaced by a per-condition decomposition.
% \begin{itemize}
%     \item \textit{Between-motion scenario:} When users transition across different motion states (e.g.\ from sitting to walking), head movements, device orientation changes, and increased screen distance collectively account for over 70\% of the variance in gaze error.
%     \item \textit{Within-motion scenario:} Even within a single motion state, micro-movements, such as the trembling while typing and the swaying of the wrist, explain significant error. Our Lasso regression shows that head distance, head movements, and orientation deflection comprise about 80\% of the regression‐weight sum, implying that small head or device tilt changes still degrade accuracy.
% \end{itemize}

\begin{itemize}
    \item \textit{Calibration-test mismatch:} The train-test matrix shows that performance is best when calibration and testing share the same motion condition, and degrades when the calibrator is transferred to another condition. This pattern indicates that the sampled geometry of calibration is a key determinant of later accuracy.
    \item \textit{Condition-sensitive factor rankings:} The sparse regression analyses repeatedly identify head-to-camera distance, head movement, and device-orientation change as important feature groups, but the per-condition decomposition shows that their relative contribution varies. This supports condition-aware mitigation rather than a single universal correction factor.
\end{itemize}

This vulnerability does not imply that 2D methods are unsuitable for handheld use. Rather, it suggests that 2D methods need additional support when face-camera-screen geometry drifts beyond the sampled calibration regime, for example through lightweight geometry normalisation, adaptive confidence estimation, or behaviour-matched calibration.

% \subsubsection{Implications for mobile eye tracking}
% A one-off calibration is often insufficient for highly dynamic daily usage. Our data suggests that small posture shifts or task switches can degrade errors by up to 2 or 3 times. A more adaptive calibration approach to periodically re-estimates the user’s head pose or normalises on‐screen coordinates is crucial for robust performance. Incorporating orientation and head‐distance data directly into the 2D pipeline may help enhance the resilience. This practice of handle head pose explicitly is often used in 3D gaze estimation pipeline. Frequent manual calibrations can disrupt flow, leading to \textit{calibration fatigue}. The approach therefore needs to be combined with natural human-device interaction to explore \emph{implicit calibration cues}; e.g., using known UI elements where the user must look to avoid user burden issues.

\subsubsection{Implications for mobile eye tracking}
The practical implication is not simply that calibration should be ``better,'' but that it should be aware of geometry, task phase, and behavioural match. The results suggest four design directions.

\noindent\textit{Transition-triggered recalibration.}
Because task-switch periods exhibited substantially larger geometry fluctuations than stable in-task periods, transitions are plausible moments to trigger lightweight recalibration, reduce confidence, or temporarily fall back to coarser interaction techniques. Such triggers should be based on detected geometry change rather than on task labels alone.

\noindent\textit{Condition-aware confidence estimation.}
The condition-specific factor rankings indicate that uncertainty estimation should incorporate geometry-related signals such as head-to-camera distance, head pose, and device orientation, rather than relying only on visual features from the face image. The relevant confidence features may differ across usage contexts; for example, walking conditions may benefit more from head-motion indicators, while standing may require stronger distance and orientation monitoring.

\noindent\textit{Behaviour-matched calibration.}
The gap between one-off, task-specific, and motion-specific calibration suggests that calibration quality depends on how well the calibration regime matches subsequent usage. For systems deployed in handheld contexts, a short seated calibration may be inadequate if the intended usage includes walking, posture switching, or frequent device reorientation.

\noindent\textit{Geometry-driven rather than time-only adaptation.}
The temporal-drift analysis did not show systematic degradation across a single 9-point test sequence. This suggests that, at least at the time scale studied here, recalibration policies should not rely on elapsed time alone. Geometry change, task transition, and calibration-test mismatch appear to be more informative triggers.

These implications are provisional and derived from a small semi-controlled study, but they provide more actionable guidance than the generic statement that motion degrades performance.

\subsection{Limitations}

Several limitations should shape the interpretation of the findings.

\noindent\textit{Small and demographically narrow sample.}
The participant pool (\(n=10\)) is a limitation, particularly because participants were drawn from a narrow university population. We therefore do not present the sample as sufficient for stable population-level estimates. The study is better understood as a dense within-subject exploratory evaluation: the same participants completed all conditions and contributed large amounts of synchronised IMU, camera, calibration, and gaze-error data. This design supports condition-sensitive analysis within the present protocol, while broader demographic and behavioural variation should be tested in larger follow-up studies.

\noindent\textit{Semi-controlled rather than unconstrained use.}
Although the tasks were designed to resemble everyday mobile activities, both studies were conducted in a laboratory environment. User Study~2 in particular required explicit calibration and 9-point testing, which likely reduced motion variance relative to more natural handheld behaviour. As a result, Study~2 should be interpreted as a constrained deployment study rather than a direct measure of unconstrained mobile usage.

\noindent\textit{Behavioural suppression during evaluation.}
A central methodological trade-off in this work is that accurate gaze-error measurement requires structured test procedures, but those same procedures can suppress the very motion variation one may wish to study. We therefore interpret User Study~1 as a reference for broader handheld geometry variation, and User Study~2 as a controlled estimate of how a calibration-based RGB 2D pipeline behaves when evaluated under task-elicited but partially constrained motion.

\noindent\textit{Instrumented phone use in User Study~1.}
In User Study~1, the experimental device was attached to the participant's own phone to preserve familiar apps and interaction habits while protecting privacy. This choice also increased device weight and thickness, which may have affected holding posture and motion. The behavioural patterns should therefore be interpreted as everyday-style but still instrumented phone use.

\noindent\textit{Asymmetry between the two studies.}
User Study~1 contained four motion conditions, whereas User Study~2 included an additional standing condition. This asymmetry was introduced to keep the observational protocol feasible while still separating upright static posture from locomotion in the evaluation study. Nevertheless, it weakens direct cross-study comparison and should be kept in mind when interpreting the results.

\noindent\textit{Descriptive factor rankings.}
The regression analysis identifies sensor-feature groups associated with error, but it should not be read as a causal decomposition of gaze-estimation failure. The dataset is small relative to the feature dimensionality, several motion signals are correlated, and the per-condition Lasso profiles are descriptive. We therefore use the factor rankings as design-oriented indicators of where mitigation may be useful, rather than as fixed properties of mobile use.

\noindent\textit{Representative pipeline rather than exhaustive model comparison.}
We used a practically relevant 2D front-camera pipeline as an empirical probe, but we did not compare a broad range of 2D and 3D gaze-estimation architectures under the same protocol. Future work should test whether explicit 3D normalisation, geometry-aware 2D methods, or hybrid approaches reduce the condition-specific failure modes identified here.

\subsection{Future directions for handheld 2D gaze estimation}

\noindent\textit{Continuous and implicit calibration.}
A single brief calibration is unlikely to expose the model to sufficient user-device geometry variation for robust handheld deployment. Future work should therefore explore incremental or implicit calibration strategies that acquire supervision during use, for example at moments of task transition, known UI fixation, or detected posture shift. The temporal-drift result suggests that such methods should be triggered by geometry change rather than by elapsed time alone.

\noindent\textit{Geometry-aware uncertainty estimation.}
The factor analyses indicate that motion-related geometry signals contain information about when the gaze-estimation pipeline is likely to fail. A promising next step is therefore to build confidence models that combine image-based features with head-to-camera distance, head pose, and device-orientation signals, while allowing the weighting of these signals to vary by motion condition.

\noindent\textit{2D--3D hybrid modelling.}
Our results suggest that the main unmet need in handheld 2D gaze estimation is not necessarily a fully new input modality, but better handling of geometry drift. A practical direction is to combine the efficiency of 2D pipelines with lightweight geometric normalisation, so that large pose and orientation changes are partially corrected without incurring the full cost of a conventional 3D pipeline.

\section{Conclusion}

% This study provides empirical evidence on the limitations of 2D gaze estimation methods in mobile contexts, particularly their vulnerability to head pose and device orientation variations. By combining motion (IMU) and vision data, we demonstrate that 2D gaze estimators struggle to handle the frequent orientation changes characteristic of real-world mobile device usage. Through two comprehensive user studies investigating the spatial dynamics of user-device interactions, we show that the assumption that models can implicitly learn to compensate for head pose variations proves inadequate in dynamic mobile scenarios. These findings highlight the need for more advanced, adaptive eye-tracking systems that can explicitly account for head movements and maintain accuracy across diverse usage contexts.  Future research should focus on developing novel approaches that address these limitations, potentially by incorporating elements of 3D gaze estimation or explicit head pose modelling into more efficient 2D frameworks that explore 2D-3D hybrid pipelines that balance real-time performance with energy efficiency.

This paper examined how motion-related geometry change affects a representative front-camera RGB 2D gaze-estimation pipeline during handheld mobile interaction. Across two semi-controlled within-subject studies, the findings show that the challenge is structured rather than simply a uniform effect of mobility. User-device geometry was comparatively stable during sustained in-task interaction, but became substantially more variable during task switches. In the gaze-evaluation study, one-off calibration improved the base model but did not fully cover the geometry encountered during later use, while task-specific and motion-specific calibration reduced error when calibration and test behaviour were better matched.

The factor analyses further indicate that head-to-camera distance, head movement, and device-orientation change are repeatedly associated with error, with their relative importance varying by motion condition. These results should be interpreted within the scope of the small, semi-controlled study, but they provide design-oriented evidence for motion-aware calibration, geometry-aware confidence estimation, and lightweight 2D--3D hybrid modelling. The absence of systematic within-sequence temporal drift further suggests that adaptation should be triggered by detected geometry change and task transitions, rather than elapsed time alone.

\appendix
\section{Supplementary regression model comparison}\label{app:regression}
This appendix accompanies the factor-identification analysis in Section~\ref{sec:results}. Its purpose is to clarify how the supplementary regression comparison was constructed and how it should be interpreted. To check whether the factor-level interpretation depends on the choice of Lasso, we compared five regression models trained to predict the mean RMSE for each participant-by-motion-condition cell from the corresponding averaged motion-feature vector, using \(n=50\) cells and 477 standardised features after removing zero-variance columns. Five-fold cross-validated $R^2$ and mean absolute error (MAE) are summarised in Table~\ref{tab:regression-comparison}.

% \textcolor{rev2}{ The comparison is used to check whether the broad factor-level interpretation depends on Lasso rather than to select a deployment model. OLS, Ridge, and Random Forest retained head-to-camera distance, orientation-related features, and head movement among their top groups, while the more heavily regularised linear models retained a sparser subset. The appendix reports cross-validated predictive performance ($R^2$, MAE) and the overlap of the top-ranked feature groups across models.}

\begin{table}[!htbp]
\centering
\small
\begin{tabular}{l|ccc}
\hline
\textbf{Model} & \textbf{ $R^2$} & \textbf{ MAE (cm)} & \textbf{Top factor groups} \\
\hline
Ordinary Least Squares & $0.32 \pm 0.29$ & $0.66 \pm 0.20$ & distance, orientation, head movement \\
Ridge                  & $0.32 \pm 0.29$ & $0.66 \pm 0.20$ & distance, orientation, head movement \\
Elastic Net            & $-0.07 \pm 0.15$ & $0.85 \pm 0.13$ & orientation, head movement \\
Random Forest          & $0.16 \pm 0.29$ & $0.74 \pm 0.12$ & distance, orientation, head movement \\
Lasso                  & $-0.09 \pm 0.21$ & $0.87 \pm 0.16$ & head movement, orientation \\
\hline
\end{tabular}
\caption{Supplementary comparison of regression models used for factor identification, evaluated by 5-fold cross-validation on participant-by-motion-condition aggregated features (\(n=50\) cells, 477 standardised features). OLS, Ridge, and Random Forest recover broadly similar geometry-related groups; Lasso is retained for the main analysis because it yields the sparsest coefficient profile.}
\label{tab:regression-comparison}
\end{table}

The two unpenalised and lightly penalised linear models (OLS and Ridge with $\alpha = 1.0$) achieved comparable cross-validated $R^2 \approx 0.32$ and MAE $\approx 0.66$\,cm, while the more heavily regularised linear models and Random Forest performed worse on this small, high-dimensional dataset, with $R^2$ between $-0.09$ and $0.16$. 
% We therefore do not use this comparison to claim that Lasso is the best predictive model. Instead, 
The comparison serves as a sensitivity check for the broad factor groups: head-to-camera distance, device-orientation/magnetometer features, and head movement recur among the highest-ranked groups in OLS, Ridge, and Random Forest, while the more heavily regularised linear models retain a sparser subset of this pattern. Lasso is retained for the descriptive factor decomposition because it provides the sparsest coefficient profile; the rankings should therefore be read as design-oriented indicators rather than stable population-level coefficients.

\section{Per-condition factor decomposition: methodological notes}\label{app:per-condition}
This appendix complements the per-condition factor decomposition reported in Table~\ref{tab:factor-by-condition}. The reported percentages represent the share of $\sum |\beta|$ attributed to each feature group within that motion condition; they should not be read as percentages of explained variance or as causal effect sizes. Values are rounded, and rows do not always sum to 100\% because small residual ``other'' features are omitted from the summary.

Because each per-condition decomposition is based on \(n=10\) participant-level observations and many correlated sensor features, the Lasso solution can become very sparse. Rows such as \emph{lying} or the two \emph{walking} conditions should therefore be interpreted as selected-feature summaries under this modelling setup, not as evidence that other geometry signals have no role. The main takeaway is the condition-sensitive pattern: different usage contexts emphasise different observable signals, suggesting that mitigation strategies may need to adapt to the geometry detected during use.

To examine whether the dominant groups differ by condition, we additionally decomposed the standardised Lasso coefficient mass into three broad groups: head-to-camera distance, head movement (head Euler angles), and device-orientation change (orientation and magnetometer features). Table~\ref{tab:factor-by-condition} summarises this descriptive decomposition. The pattern is condition-dependent: head-to-camera distance contributes most clearly in \emph{standing}, while the walking conditions are dominated by head-movement features in this sparse model. The practical implication is that mitigation may need to be condition-aware, for example prioritising depth normalisation when viewing distance changes and confidence adjustment when head-motion features dominate.

\begin{table}[!htbp]
\centering
\small
\begin{tabular}{l|ccc}
\hline
\textbf{Motion condition} & \textbf{Head distance} & \textbf{Head movement} & \textbf{Device orientation} \\
\hline
Lying            & $\approx$0\%  & $\approx$0\%   & $\approx$0\%  \\
Sitting          & $\approx$0\%  & $\approx$15\%  & $\approx$20\% \\
Standing         & $\approx$33\% & $\approx$18\%  & $\approx$43\% \\
Walking          & $\approx$0\%  & $\approx$100\% & $\approx$0\%  \\
Walking in maze  & $\approx$0\%  & $\approx$100\% & $\approx$0\%  \\
\hline
\end{tabular}
\caption{Per-condition decomposition of Lasso coefficient mass into three broad factor groups (head-to-camera distance, head movement, device-orientation change). Percentages represent the share of $\sum |\beta|$ within that condition. Methodological notes (rounding, residual ``other'' group, and the degenerate behaviour of the lying and walking conditions under the small per-condition sample size) are given in~\ref{app:per-condition}.}
\label{tab:factor-by-condition}
\end{table}

% \appendix
% \section{Example Appendix Section}


% ===== Begin inlined bibliography =====

% ===== End inlined bibliography =====

\end{document}